\documentclass[sn-aps,iicol]{sn-jnl}
\usepackage{graphicx,upgreek}
\usepackage{lipsum}
\usepackage{natbib}
\usepackage{natbib}
\usepackage{hyperref}
\hypersetup{
    colorlinks=true,
    linkcolor=blue,
    filecolor=magenta,      
    urlcolor=cyan,
    pdftitle={Overleaf Example},
    pdfpagemode=FullScreen,
    }
\usepackage{physics}
\usepackage{amsmath}
\jyear{2022}%
\theoremstyle{thmstyleone}%
\theoremstyle{thmstyletwo}%
\theoremstyle{thmstylethree}%
\usepackage{etoolbox}\apptocmd\appendix{\pretocmd\section{\clearpage}{}{}}{}{}
\begin{document}


\title[Article Title]{Heterogeneous integration of spin-photon interfaces with a scalable CMOS platform}

\author[1,2]{\fnm{Linsen} \sur{Li\textsuperscript{*,}}}
\author[1,3]{\fnm{Lorenzo} \sur{De Santis}}
\author[1,2]{\fnm{Isaac} \sur{Harris}}
\author[1,2]{\fnm{Kevin C.} \sur{Chen}}
\author[1]{\fnm{Yihuai} \sur{Gao}}
\author[1,2]{\fnm{Ian} \sur{Christen}}
\author[1,2]{\fnm{Hyeongrak} \sur{Choi}}
\author[1,2,4]{\fnm{Matthew} \sur{Trusheim}}
\author[1]{\fnm{Yixuan} \sur{Song}}
\author[1,3]{\fnm{Carlos} \sur{Errando-Herranz}}
\author[1]{\fnm{Jiahui} \sur{Du}}
\author[1,2]{\fnm{Yong} \sur{Hu}}
\author[1,5]{\fnm{Genevieve} \sur{Clark}}
\author[6]{\fnm{Mohamed I.} \sur{Ibrahim}}
\author[5]{\fnm{Gerald} \sur{Gilbert}}
\author[1,2]{\fnm{Ruonan} \sur{Han}}
\author[1,2]{\fnm{Dirk} \sur{Englund\textsuperscript{\dag,}}}
\affil[1]{\orgdiv{Research Laboratory of Electronics}, \orgname{Massachusetts Institute of Technology}, \orgaddress{ \city{Cambridge}, \postcode{02139}, \state{MA}, \country{USA}}}
\affil[2]{\orgdiv{Electrical Engineering and Computer Science}, \orgname{Massachusetts Institute of Technology}, \orgaddress{ \city{Cambridge}, \postcode{02139}, \state{MA}, \country{USA}}}
\affil[3]{\orgdiv{QuTech}, \orgname{Delft University of Technology}, PO Box 5046, 2600 GA Delft, Netherlands}
\affil[4]{\orgdiv{DEVCOM}, \orgname{Army Research Laboratory}, \orgaddress{ \city{Adelphi}, \postcode{20783}, \state{MD}, \country{USA}}}
\affil[5]{\orgname{The MITRE Corporation}, \orgaddress{\city{Bedford}, \state{MA}, \country{USA}}}
\affil[6]{\orgdiv{School of Electrical and Computer Engineering}, \orgname{Cornell University}, \orgaddress{ \city{Ithaca}, \postcode{14853}, \state{NY}, \country{USA}}}
\email{linsenli@mit.edu\textsuperscript{*}}
\email{englund@mit.edu\textsuperscript{\dag}}

\abstract{Color centers in diamonds have emerged as a leading solid-state platform for advancing quantum technologies, satisfying the DiVincenzo criteria~\cite{divincenzo2000physical} and recently achieving a quantum advantage in secret key distribution~\cite{bhaskar2020experimental}. Recent theoretical works~\cite{choi2019percolation,nickerson2014freely,nemoto2014photonic} estimate that general-purpose quantum computing using local quantum communication networks will require millions of physical qubits to encode thousands of logical qubits, which presents a substantial challenge to the hardware architecture at this scale. To address the unanswered scaling problem,  in this work, we first introduce a scalable hardware modular architecture ``Quantum System-on-Chip'' (QSoC) that features compact two-dimensional arrays ``quantum microchiplets'' (QMCs) containing tin-vacancy (SnV$^-$) spin qubits integrated on a cryogenic application-specific integrated circuit (ASIC). We demonstrate crucial architectural subcomponents, including \textbf{(1)} QSoC fabrication via a lock-and-release method for large-scale heterogeneous integration; \textbf{(2)} a high-throughput calibration of the QSoC for spin qubit spectral inhomogenous registration; \textbf{(3)} spin qubit spectral tuning functionality for inhomogenous compensation; \textbf{(4)} efficient spin-state preparation and measurement for improved spin and optical properties. QSoC architecture supports full connectivity for quantum memory arrays in a set of different resonant frequencies and offers the possibility for further scaling the number of solid-state physical qubits via larger and denser QMC arrays and optical frequency multiplexing networking. }

\maketitle
\begin{figure*}[ht]%
\centering
\includegraphics[width=1\textwidth]{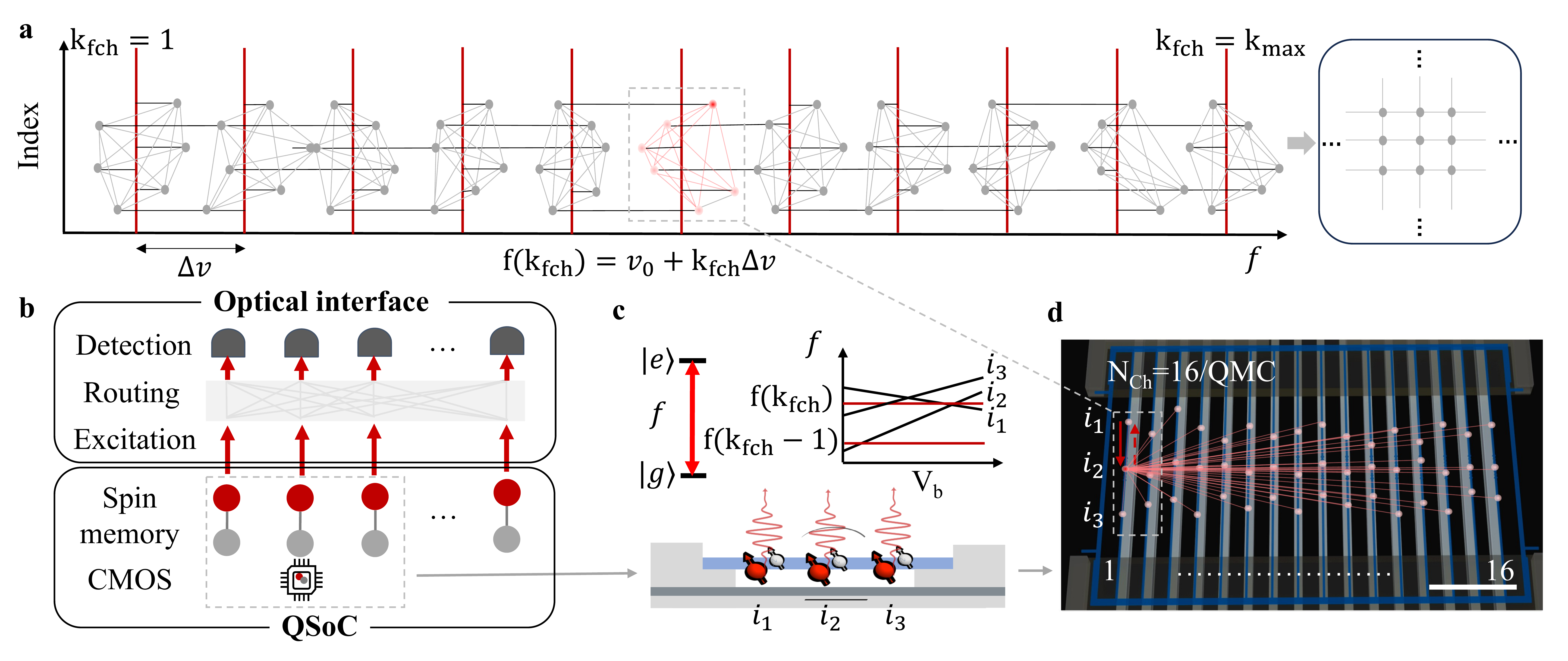}
\caption{\textbf{Comprehensive architectural design.} \textbf{a,} illustration of the architecture goal for building a connected qubit graph with qubits resonant with a set of frequencies (red lines: $f(\text{k}_{\text{fch}})=v_{0}+\text{k}_{\text{fch}}\Delta v$). The gray dot represents the artificial atom qubit, where the index indicates the $i_{\text{th}}$ qubit (marked by location) that can be tuned to a specific frequency $\text{k}_{\text{fch}}$ at the minimum, mean, or maximum tuning voltage. The horizontal black line represents the spectral tunability of this qubit. Multiple gray dots in one horizontal line mean that it is the same spin qubit at different tuning voltages, demonstrating its wide tuning range across various frequencies. The gray line between the gray dots means that those two qubits can be entangled, as they can tune to the same frequency. \textbf{b,} A comprehensive architecture diagram, illustrating the optical interface (including optical excitation, routing, and detection) and the QSoC (including the CMOS ASIC chip and spin memories). The red dots indicate the electronic spin of the qubit (quantum emitter), and the gray dots represent the corresponding nuclear spin (quantum memory). \textbf{c,} A cross-section diagram of three quantum emitters ($i_1$, $i_2$, and $i_3$) located in a waveguide. The optical transition from ground to excited state has a transition frequency $f$, which is tunable with the system tuning voltage bias V$_\text{b}$. The $f$ versus V$_\text{b}$ of different emitters is shown here indicating the different behaviors for the voltage tuning response of $f$ as a function of V$_\text{b}$ that can be applied to the emitters. Here, we can align the emitters to a set of frequencies f(k$_\text{fch}$). \textbf{d,} An expanded view illustrating quantum channels, numbered from 1 to 16, within a QMC (scale bar: 10$\upmu$m). Illustration indicates that an emitter can interact with all other emitters of the same resonant frequency through free-space optical routing and detection. The gray box region provides a practical example of the diagram in \textbf{a}, but much more resonant quantum emitters a certain quantum emitter can interact across the entire chip.} \label{fig1}
\end{figure*}



\textbf{Introduction -} Modularity plays a critical role in computing architectures, allowing the segregation and combination of diverse system components. This principle has been applied to quantum information processing, resulting in quantum networks consisting of multiple processing units interconnected through coherent channels. Such networks have been proposed for trapped ion~\cite{monroe2014large}, neutral atom~\cite{welte2018photon}, and spin-based systems~\cite{choi2019percolation,nickerson2014freely,nemoto2014photonic}, with the aim of achieving scalable distributed quantum processing. However, it is essential for the scalability of the qubit layer to be used for building a large-scale quantum system. In this study, we introduce a scalable hardware modular architecture ``Quantum System-on-Chip'' (QSoC) that leverages the fabrication of the qubit layer with modern mass microfabrication processes for scalability. The central qubit platform in our proposed architecture utilizes electron-nuclear spin systems of diamond color centers which can be generated with the foundry ion implantation process for large-scale fabrication compared with the individual fine-operated ion and atom. Diamond color centers have emerged as promising solid-state qubits, demonstrating deterministic remote entanglement~\cite{humphreys2018deterministic}, minute-long coherence times with more than ten auxiliary qubits~\cite{bradley2019ten}, and large-scale heterogeneous integration into photonic integrated circuits using diamond quantum microchiplets (QMCs)~\cite{wan2020large}. The implanted diamond color center provides benefits in scalability, but will suffer from naturally inhomogeneous spectral broadening~\cite{wan2020large}. To address that, we integrate the implanted diamond qubit layer on a commercially processed complementary metal-oxide-semiconductor (CMOS) backplane, allowing for local spectral tuning. The system's co-design with CMOS electronics allows the compact two-dimensional array of qubit arrangements and substantially minimizes the size of the system's control elements. CMOS electronics have been used in gate-defined quantum dot control~\cite{xue2021cmos}, low-power cryogenic microwave control~\cite{bardin2019design}, and integration with nitrogen vacancy (NV) centers for sensing applications~\cite{kim2019cmos,ibrahim2020high}. 

The QSoC demonstrated here contains 64 QMCs with a co-designed CMOS application-specific integrated circuit (ASIC). QSoC not only facilitates qubit scaling but also enables qubit inhomogeneous compensation for full connectivity, which benefits the cluster state computational power that is related to the size of the largest connected qubit graph size~\cite{choi2019percolation}. The connected qubit graph can be mapped to a regular qubit square lattice layout in the quantum circuit representation for error correction~\cite{choi2019percolation}. In this paper, we show the architectural design concept first, followed by the QSoC fabrication, characterization for spin qubit spectral inhomogenous registration, and spin qubit tuning for inhomogeneous compensation to demonstrate the QSoC capability can achieve the architecture required function. Here we report an unprecedented scale comprising over 10,000 individual resolved diamond spin-photon interfaces in the QSoC architecture designed for the rapid generation of fully connected qubit graphs. We also demonstrate that we can maintain the quantum emitter's spin and optical properties with efficient spin-state preparation and measurement on such a novel platform and analyze the QSoC tunability requirement as we further scale up the system size in the future.

\begin{figure*}[ht]%
\centering
\includegraphics[width=1\textwidth]{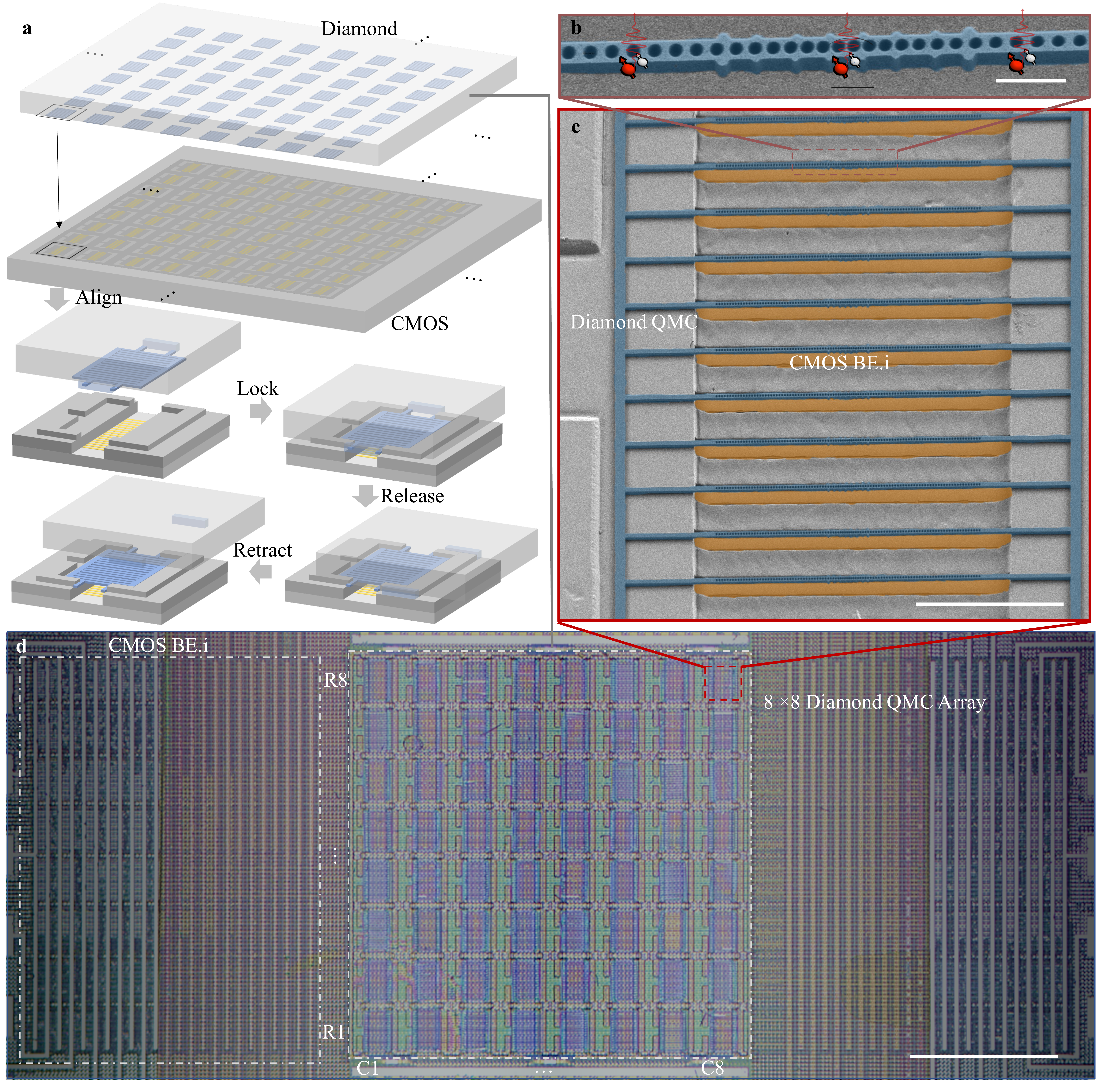}
\caption{\textbf{Fabrication.} \textbf{a,} Schematic representation of the lock-and-release heterogeneous integration process employed to transfer the quantum microchiplet (QMC) array, which is heterogeneously integrated on the post-processed CMOS chip, to the diamond parent chip. This process involves alignment, locking, releasing, and retracting. \textbf{b,} A scanning electron microscope (SEM) image of a quantum channel (light blue) within the QMC, utilizing a dielectric antenna design optimized for photon collection in free space (scale bar 1$\upmu$m). Some quantum emitter may be coupled to the antenna like the central one here. \textbf{c,} An enlarged SEM image of \textbf{d} highlighting the central region of a single QMC (scale bar 10$\upmu$m). The orange color indicates the visible CMOS individual backplane electrodes (BE.i). \textbf{d,} An optical microscope image of 1024 diamond resonant dielectric antennas integrated on the CMOS control chip, providing a broader view of the QMC array with rows 1-8 and columns 1-8 (scale bar 200$\upmu$m).}\label{fig2}
\end{figure*}

\textbf{Architecture goal for building a fully connected qubit graph -} 
The largest connected qubit graph determines the quantum computational power of the cluster state~\cite{choi2019percolation} so we would like to build a fully connected qubit graph to utilize as much as qubit resource we have in the system. Figure~\ref{fig1}a illustrates the goal of creating an interconnected qubit graph, where each qubit resonates within a specific set of frequencies channel, which number k$_\text{fch}$ labeled from 1 to k$_\text{max}$ for herald entanglement. This is due to the wide spectral range of ZPL in the implanted diamond color centers. We show an example of using the uniform frequency channel distribution with spacing $\Delta v$ here. During the pre-characterization phase, we get a lookup table for each frequency. This table indexes various quantum emitters, detailing their positions and the corresponding voltages needed to resonate within a particular frequency channel.

In each frequency channel depicted, a few indexed quantum emitters are shown as examples. A red line indicates a resonant frequency channel we aim to match by tuning the quantum emitter ZPL with the CMOS backplane. Each gray dot in the figure represents a qubit, and the horizontal black line shows its potential tuning range by the CMOS backplane. A qubit that can be tuned to match a specific frequency channel f(k$_\text{fch}$) is indexed in the lookup table for that channel. We consider two quantum emitters interference-able~\cite{bernien2013heralded} when they align with the same frequency channel, and their emitted ZPL photons, sharing identical polarization after filtering, fall within twice the transform-limited linewidth. This is visually represented by a pale gray line connecting the dots.

Consequently, all the gray dots form a large, interconnected graph, allowing for communication between every qubit. This graph is dynamically reconfigurable to suit various task requirements, including potential cluster connections. For example, the qubit graph is able to reprogrammed into a surface code grid structure to support error correction algorithms~\cite{choi2019percolation}. Moreover, by selectively using high-quality quantum emitters with narrow linewidths and bright spin-photon interfaces, this architecture is made robust against fabrication variations in the hardware.



\textbf{Comprehensive system architecture -} Figure~\ref{fig1}b presents a comprehensive architecture diagram that includes the optical interface and QSoC. The optical interface encompasses optical excitation, routing, and detection (see Appendix C). The QSoC consists of a CMOS ASIC cooled to 4~K in a cryostat, which is heterogeneously integrated with a 64 diamond QMC array. The spin qubit chosen for this demonstration in QMC is SnV$^-$, offering high quantum efficiency~\cite{iwasaki2017tin} and spin performance that is compatible with cryogenic temperatures above 1~K~\cite{trusheim2020transform}. 


\textbf{QSoC module detail -} Figure~\ref{fig1}c illustrates the basic function of the QSoC module. The ASIC provides a voltage bias $V_\text{b}$ to tune the zero-phonon line (ZPL) transition frequency $f$ of the quantum emitters. These can be tuned to a predefined set of frequency channels labeled as f(k$_\text{fch}$). An example cross section in the QSoC showcases the tuning response behavior of different quantum emitters ($i_1$, $i_2$, and $i_3$) with varying $f$ as a function of V$_\text{b}$. Some quantum emitters, such as $i_2$, can couple to a resonant dielectric antenna to enhance free space coupling (see Methods). 


Figure~\ref{fig1} d shows a 3D representation of the CMOS circuit layout co-integrated with diamond QMCs, each featuring N$_\text{ch}$~=~16 channels. Each quantum channel is integrated with the diamond-resonant dielectric antenna that designs an efficient optical interface based on a 1D photonic crystal cavity design. The antenna incorporates a cointegrated vertically radiating grating coupler, resulting in a greater efficiency of 96\% free space collection efficiency within NA~=~0.9 in simulation (see Methods for details). A metal layer on the CMOS backplane chip facilitates electronic signal routing from the external electronic source to each QMC. Future iterations of the CMOS chip can incorporate built-in digital logic and analog pulse sequence for routing of quantum control signals with external sources~\cite{pauka2021cryogenic}.

\begin{figure*}[ht]%
\centering
\includegraphics[width=0.8\textwidth]{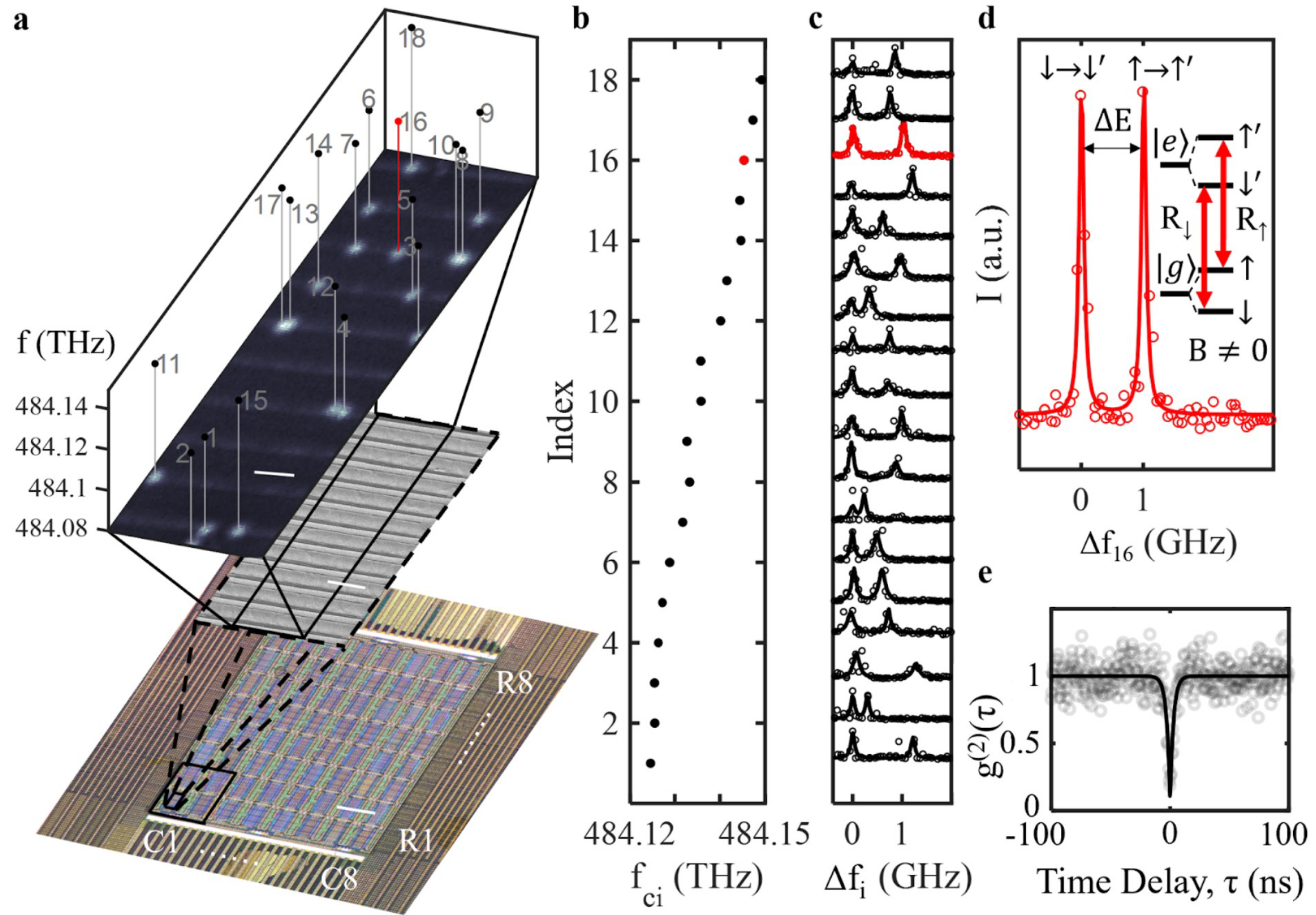}
\caption{\textbf{QSoC characterization.} \textbf{a,} An overlay of the optical microscope image (scale bar 100~$\upmu$m), SEM image (scale bar 5~$\upmu$m), and superposition EMCCD image of the emitters' bright frames (scale bar 3~$\upmu$m). These images show an optically bright SnV$^-$ under a resonant laser excitation frequency ranging from 484.123 to 484.153~THz, with corresponding index mappings indicated. The location of this example on the CMOS chip is also marked. \textbf{b,} The ZPL frequency f$_\text{ci}$ of the emitter with index $i$ in \textbf{a}. The indices are sorted according to the peak ZPL frequency from low to high, where f$_\text{ci}$ consistently represents the lowest frequency value of the double peaks in the PLE spectrum. \textbf{c,} The PLE spectrum of the marked emitters in \textbf{a} with ZPL frequency shifted by f$_\text{ci}$. \textbf{d,} An example PLE spectra of SnV$^-$ for emitter $i$ = 16 in \textbf{c}. Two spin-state transitions correspond to two peaks in the PLE with splitting $\Delta$E. The inset illustrates the energy diagram of the SnV$^-$ with and without the magnetic field. \textbf{e,} Autocorrelation measurements of the single SnV$^-$ in \textbf{d}.}\label{fig3}
\end{figure*}

\textbf{Fabrication with lock-and-release integration -} We introduced SnV$^-$ centers through ion implantation and high temperature annealing in diamond, followed by QMC nanofabrication to define the transferable diamond nanostructure on the bulk parent diamond surface (see Methods). The lock-and-release scalable heterogeneous integration technology, a crucial step in QSoC fabrication, is illustrated in Fig.~\ref{fig2}a (the detail of the cross section in Fig.~\ref{figS3}b). This procedure enables the parallel transfer of a quantum memory matrix consisting of 8 columns (C1-C8) and 8 rows (R1-R8) to the central region (500~$\upmu$m $\times $ 500~$\upmu$m) of the CMOS chip socket, which includes N$_\text{sys}=1024$ of quantum channels in total (see Methods for details). Here, we flip the diamond parent chip and align it with a locking structure that has been post-fabricated on the TSMC 180~nm CMOS chip, followed by fine adjustment of the QMCs with two probes (see Methods and Appendix A). After alignment, we move the parent bulk diamond vertically to lock the QMC and then horizontally, as depicted in Fig.~\ref{fig2}a, to break the bridges between the QMC and the bulk diamond for release, followed by retraction of the bulk diamond. Figure~\ref{fig2}b shows an SEM image of a single central quantum channel. Figure~\ref{fig2}c presents an SEM image of a single QMC region of the chip, with the orange region indicating the individual CMOS backplane electrode region beneath the QMCs. Figure~\ref{fig2}d displays an optical microscope image of the 1024 quantum channels integrated into the CMOS control chip. For each quantum channel, we expect to have around 3 resonant quantum emitters on average at a certain optical frequency (see the discussion later). The number of quantum channels in this design can be readily scaled by using a larger CMOS chip corresponding to the diamond parent chip size.



\begin{figure*}[ht]%
\centering
\includegraphics[width=1\textwidth]{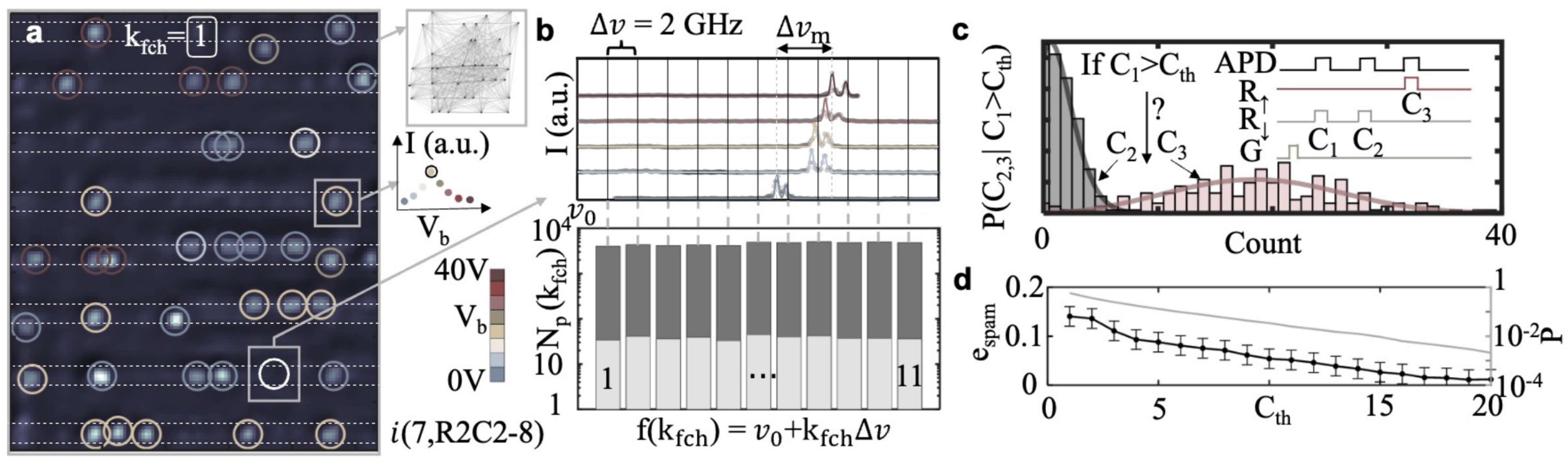}
\caption{\textbf{Quantum emitter spectral tuning and spin state preparation and measurement.} \textbf{a,} Widefield images of the QMC region at k$_\text{fch}$ = 1. Each emitter is labeled with a circle whose color indicates the voltage tuning $V_\text{b}$ at which it is brightest. \textbf{b,} An example of voltage-induced tuning for an emitter crossing various frequency channels. The statistics of the emitter bright spots number are shown underneath for 11 frequency channels connected by the lines, representing both shallow gray (in one field of view) and deep gray (among all the targeted 1024 quantum channels). \textbf{c,} The spin state preparation and measurement pulse sequence, involving four programmable channels: green repump laser (G), resonant laser 1 (R$_\downarrow$)resonating with the lower frequency transition ($\ket{\downarrow}$ to $\ket{\downarrow}'$), resonant laser 2 (R$_\uparrow$) resonating with the other higher frequency transition ($\ket{\uparrow}$ to $\ket{\uparrow}'$), and APD. We have three collection time bins (each of duration T$_\text{M}$). The APD time bin 1 serves as the spin state preparation signal for post-selection. We show histogram measurement counts, post-selected with state preparation threshold counts C$_\text{th}$=18 (APD time-bin 1 readout) with T$_\text{M}$=~50~$\upmu$s. Following post-selection, APD time-bin 2 measures the dark state count, while APD time-bin 3 measures the bright state counts. \textbf{d,} The relationship between state preparation and measurement error e$_\text{spam}$ (black line with left axis) and successful post-selection probabilities, p (gray line with right axis) with C$_\text{th}$.}\label{fig4}
\end{figure*}



\textbf{System parameters -} In the following sections, we discuss the measurement of essential parameters of the QSoC to estimate our scaling benefit while keeping the diamond color center optical and spin performance in the QSoC. Essential parameters include \textbf{(1)} System size N$_\text{sys}$, determined by the target number of quantum channels achievable through lock-and-release heterogeneous integration; \textbf{(2)} Emitter number per quantum channel n$_\text{emitter}$, representing the maximum proportion of the quantum emitter in each quantum channel that can be tuned to the same ZPL frequency (see Methods); \textbf{(3)} Spin qubit state preparation and measurement error e$_\text{spam}$; \textbf{(4)} The spin-photon interface efficiency, characterized by the coherent photon detection probability p$_\text{det}$ after spin state initialization. We also consider the potential nanocavity enhancement (Purcell factor F$_\text{p}$) of a resonant dielectric antenna, which can boost the ZPL photon collection rate. 


\textbf{Scalable SnV$^-$ characterization -} We perform a high-throughput characterization of our SnV$^-$ qubits using optical excitation with wide-field illumination and readout from an electron-multiplying charge-coupled device (EMCCD)~\cite{sutula2023large} (see Methods). We show an example of this measurement on a QMC in Fig.~\ref{fig3}. Figure~\ref{fig3}a presents the spatial locations of SnV$^-$ in the central QMC region. Figure~\ref{fig3}b shows the ZPL frequency of the color centers f$_\text{ci}$, and their normalized photoluminescence excitation (PLE) spectra are reported in Figure~\ref{fig3}c. All measurements and discussions regarding the transition between the ground state and excited state of SnV- throughout the manuscript refer to the ZPL C line. Figure~\ref{fig3}d shows an example PLE of an SnV$^-$ with an external magnetic field B $=$ 0.13~T along the diamond axis [001], revealing the two spin-conserving transitions utilized for spin initialization and readout. The spin state can be controlled by an external microwave signal~\cite{trusheim2020transform, golter2022multiplexed} or a modulated laser~\cite{debroux2021quantum}. The ZPL frequency of SnV$^-$ can be tuned to overcome the inhomogeneous distribution via strain~\cite{wan2020large,aghaeimeibodi2021electrical}. To confirm the presence of a single quantum emitter, we performed a second-order autocorrelation (g$^{(2)}$) measurement using resonant laser excitation and PSB collection. The collected light is split into two avalanche photodiode (APD) collection paths. The g$^{(2)}$ measurement of SnV$^-$ in Fig.~\ref{fig3}d is illustrated in Fig.~\ref{fig3}e. Here, g$^{(2)}$(0) is 0.07 without background correction, indicating the presence of a single emitter as g$^{(2)}$(0)$<$0.5.






\textbf{SnV$^-$ spectral tuning -} Individual SnV$^-$ can typically be spectrally resolved due to the inherent inhomogeneous distribution of their optical transitions~\cite{bersin2019individual}. The SnV$^-$ optical frequency can be tuned by means of a capacitive actuator controlled by voltage ~\cite{wan2020large,meesala2018strain} (detailed in Figure~\ref{figS7}). Based on the statistical result in Appendix B, we estimate that the average tuning range is around 2~GHz within the applicable voltage range. Here, we chose a set of uniform 11 frequency channels with spacing $\Delta v$ = 2~GHz to utilize inhomogeneously distributed quantum emitters within the mode-hop laser tuning range (see Appendix C). The QSoC can tune a quantum emitter's ZPL within the laser tuning range into one of the resonant frequency channels set on average. Figure~\ref{fig4}a displays a wide-field image of the 8 quantum channels of a QMC, focused on k$_\text{fch}$=1 frequency channels. This is superimposed with the anticipated edges of the diamond nanobeam. The image highlights bright emitters, each encircled in a color that corresponds to the optimal tuning voltage to align it with the resonant wavelength. As detailed in the inset, the process involves sweeping the voltage from 0V to 40V on the CMOS backplane to identify the voltage level that maximizes the emitter's intensity. Within the Field of View (FOV), the emitter data is represented as a point cloud shown in the inset. Here, each black dot marks the location of a quantum emitter, while the connecting gray lines indicate that these emitters can be tuned to resonate with the same frequency channel. Figure~\ref{fig4}b illustrates the tuning of an emitter from frequency channel k$_\text{fch}$ = 7 to k$_\text{fch}$ = 8 at varying voltages. The spectral tuning range from the min tuning voltage to the max tuning voltage, defined as $\Delta v_\text{m}$, is shown in the figure. Appendix B includes a simulation of the spectral tuning effect from voltage-induced strain and statistics on the emitter spectrum tuning range. The histograms underneath represent the number of emitters found in each frequency channel with light gray for a single field of view (FOV) and dark gray for all 1024 quantum channels measured in the entire CMOS chip. The number of bright spots in a single FOV at a specific frequency corresponds to the potential direct connections per qubit enabled by the all-to-all optical connection (see Appendix C). We estimate that we have approximately 2400 resonant emitters in the whole central region of the CMOS chip socket to calculate n$_\text{emitter}$ at a specific laser frequency with a maximum of 40~V CMOS backplane tuning.



\textbf{Spin state preparation and measurement -} Figure~\ref{fig4}c demonstrates spin state preparation and measurement of the SnV$^-$ presented in Fig.~\ref{fig3}d. We collect the phonon sideband (PSB) emissions confocally using an APD. The inset reveals the pulse time sequence used for state preparation and measurement (see Methods). Initially, a green laser resets the SnV$^-$ to the negative charge state~\cite{gorlitz2022coherence}. Then, we herald the correct spin state of SnV$^{-}$ using a laser pulse resonant with the R$_\downarrow$ transition. We set a signal threshold count C$_\text{th}$, and when the APD time-bin 1 counts exceed C$_\text{th}$, we consider the SnV$^-$ spin state to be successfully prepared. We illustrate the count histogram of C2 and C3 (inset of Fig.~\ref{fig4}c) with post-selection on successful state preparation events. Based on the histogram result, we can calculate the spin state preparation and measurement error e$_\text{spam}$ (see Method). Figure~\ref{fig4}d reveals the relationship between e$_\text{spam}$ and C$_\text{th}$ on the black line, with the probabilities of successful events for different C$_\text{th}$ values by the gray line on the logarithmic scale. By selecting C$_\text{th}$ = 18 as an example, we can reduce e$_\text{spam}$ to 3\% after post-selection. Although the probability of successful initialization is about 3\% in this case, initialization can be attempted multiple times until it is successful. We expect the spin state preparation will have a successful event on average within 2~ms. The statistical result without post-selection is displayed in Appendix B.


\begin{figure*}[ht]%
\centering
\includegraphics[width=1\textwidth]{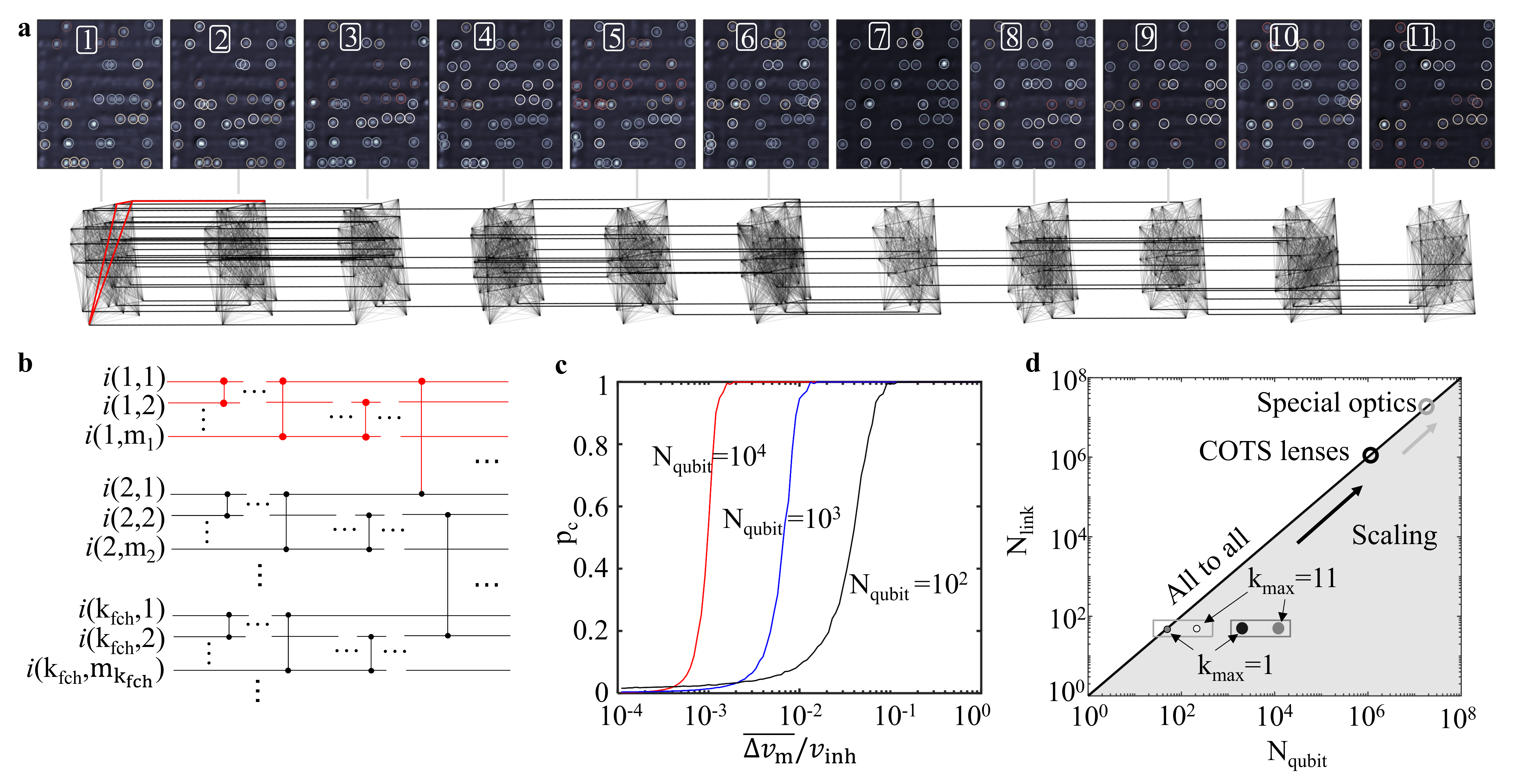}
\caption{\textbf{QSoC enables large-scale fully connected qubit graph with further scaling.} \textbf{a,} Widefield images of the emitter spot region across 11 frequency channels (k$_\text{fch}$ from 1 to 11) as well as the illustration of connected cluster using the widefield data from a single field of view.\textbf{b,} The quantum circuit representations of the connected qubit graph. \textbf{c,} The ratio of the fully connected qubit graph p$_\text{c}$ against $\overline{\Delta v_\text{m}}$/$ v_\text{inh}$ under various N$_\text{qubit}$. \textbf{d,} The scaling potential of the QSoC platform considering N$_\text{qubit}$ and N$_\text{link}$. The shading indicates the potential scaling of the corresponding system with the all-to-all connectivity line.}\label{fig5}
\end{figure*}


\textbf{Efficient spin-photon interface -} A dielectric antenna structure optimizes the photon emission of the quantum emitter for efficient free space collection. Based on the average PSB photon readout counts from the bright state of the SnV$^-$ in Fig.~\ref{fig4}b, we estimate that the lower bound of p$_\text{det}$ is above $2.4\times10^{-3}$ (see Methods), surpassing the previously reported p$_\text{det}$ of 4$\times$10$^{-4}$ ~\cite{bernien2013heralded,humphreys2018deterministic,pompili2021realization}. This indicates that our potential entanglement generation rate for an emitter pair can be several times higher than the previously reported value. The incorporation of a nanocavity Purcell effect by the resonant dielectric antenna can further enhance the ZPL photon emission rate. With an extract Purcell factor of F$_\text{p}=2.9$ for a typical SnV$^-$, the potential improvement in the ZPL photon collection rate is significant through such an efficient spin-photon interface (see Appendix B and Methods). Enhanced fabrication and alignment of the location and orientation of the emitter can further improve the Purcell factor, as simulations have shown a quality factor ten times higher compared to current devices~\cite{aghaeimeibodi2021nanophotonic,kuruma2021coupling}.  




\textbf{QSoC enables large scale fully connected qubit graph -} 
Figure~\ref{fig5}a illustrates how a connected qubit graph can be built using experimental single FOV data at 11 frequency channels (k$_\text{fch}$ from 1 to 11). In Figure 5a, an expanded version of Figure 4a is presented, encompassing all 11 frequency channels. Similar to the previous figure, bright emitters are encircled, with the circle's color denoting the tuning voltage best suited for aligning the emitter with its resonant wavelength. To determine the positions of quantum emitters, we assess whether the centers of two quantum emitters, within their respective frequency channels, are within one pixel of each other in the wide-field characterization. Additionally, if their brightness surpasses a certain threshold and the difference in brightness is less than 50\%, we group them as the same quantum emitters at varying backplane voltages. Emitters meeting these criteria are tunable across multiple frequency channels, and their connections are marked in the graph. In a FOV, the emitters have all-to-all connectivity within frequency channels with optical routing~\cite{christen2022integrated,kim20031100, palm2023modular}, and some of the emitters can be tuned to connect the neighbor frequency channels. Figure~\ref{fig5}b presents the quantum circuit of equivalence with the cluster in Fig.~\ref{fig5}a, assuming each frequency channel k$_\text{fch}$, it has m$_{\text{k}_\text{fch}}$ quantum emitters. A desired quantum algorithm can be compiled for the connectivity of the system. The red line labeled here corresponds to the red line in Fig.~\ref{fig5}a here. Figure~\ref{fig5}c indicates the ratio of the fully connected qubit graph to the total number of qubit nodes (p$_\text{c}$) as a function of the average frequency tunability ratio ($\overline{\Delta v_\text{m}}$/$v_\text{inh}$) across the entire inhomogeneous range ($v_\text{inh}$) under different sizes of the qubit system (N$_\text{qubit}$) (see Methods). This plot shows that for larger system scales, the tunability requirement for achieving a fully connected qubit graph is reduced, allowing the system to operate with lower tuning voltages for better energy efficiency. Fully connected qubit graph allows for the implementation of any error correction code with finite degree and weight. Especially, our optical reconfigurability allows for non-planar operation, opening the possibility of using recently developed codes like hyperproduct code~\cite{tillich2013quantum} or other quantum low-density parity-checking codes~\cite{breuckmann2021quantum} with a favorable encoding rate. For practical applications, we consider the surface code~\cite{kitaev1997quantum,choi2019percolation,fowler2012towards} (shown in Figure~\ref{fig1}a) for their high error threshold and optimized performances in decoding and magic state distillation~\cite{litinski2019magic}, as well as hashing-bound saturation in handling biased errors~\cite{bonilla2021xzzx}.

\textbf{Further scaling -} The QSoC module presented here highlighted the advantages of scalability in terms of qubit numbers and connectivity. Connectivity refers to the number of distinct qubits that a single qubit can interact with in an entanglement trial. Figure~\ref{fig5}d illustrates the scaling potential of the QSoC platform, considering the number of qubits (N$_\text{qubit}$) and the number of direct connections per qubit (N$_\text{link}$). The black line, N$_\text{link}$ = N$_\text{qubit}$, corresponds to all-to-all connectivity, and the shaded region below represents the physically accessible region on the hardware. The light gray box, including the smaller scattered points, represents the data from a single FOV, while the dark gray box, including the larger scattered points, represents the data from the entire sample area. Two gray boxes include the results of a single frequency channel (k$_\text{max}$ = 1) and 11 frequency channels (k$_\text{max}$ = 11).

To achieve further scaling, a larger field-of-view objective can be used (e.g. a commercial 10$\times$ Olympus plan objective with 0.3 NA and a 2.65~mm diameter FOV). Coupled with a denser QMC design (2.52 $\upmu$m spacing for diffraction-limited spots at SnV$^-$ ZPL with the given objective NA), this imaging system achieves up to 8.7 $\times 10^5$ directly resolvable spots using commercial-off-the-shelf (COTS) lenses, as shown in Fig.~\ref{fig5}. When a custom-designed lens is employed, the number of diffraction-limited spot sites can be expanded to over ten million, with each spot capable of hosting hundreds of quantum emitters. This level of qubit density of the QSoC is close to the transistor density of the most advanced semiconductor processes. (3$\times$10$^8$ transistors/mm$^2$ in TSMC N3 process~\cite{wu20223nm}). With a broader inhomogeneous distribution range, our qubit density will not be limited by the transistor density, although the transistor density would determine the number of individual voltages the system can apply simultaneously.

\begin{table}[h]
\begin{center}
\begin{minipage}{0.455\textwidth}
\caption{QSoC parameters.}\label{tab1}%
\begin{tabular*}{\textwidth}{ |p{1.8cm}|p{2.2cm}|p{2cm}| }
\hline
Parameter & Reported work & This work \\
\hline
N$_\text{sys}$ & 128~\cite{wan2020large} & 1024 \\
n$_\text{emitter}$ & 1~\cite{wan2020large} & 2.3 \\
e$_\text{spam}$ (SnV) & 25\%~\cite{gorlitz2022coherence}   & 3\% \\
p$_\text{det}$ (F$_\text{p}$) & 4$\times$10$^{-4}$~\cite{humphreys2018deterministic} (1) & 2.4$\times$10$^{-3}$ (2.9) \\
\hline
\end{tabular*}
\end{minipage}
\end{center}
\end{table}

\textbf{Conclusion -} In summary, we have demonstrated the fabrication and characterization of the core QSoC module through a scalable transfer method for high-yield, large-scale integration of artificial atom arrays with CMOS chips, along with a high-throughput characterization approach. The summarized parameters of the characterized QSoC module are presented in Table 1.

The QSoC module exhibits significant advantages in terms of the parameters N$_\text{sys}$ and n$_\text{emitter}$ parameters. Other parameters, such as e$_\text{spam}$, p$_\text{det}$, and F$_\text{p}$ can be further improved through refined material processes and control sequences. The architecture can be expanded by incorporating a superconducting nanowire for efficient single-photon detection~\cite{zhu2016superconducting}, large-scale CMOS chip control for low-latency solid-state spin control~\cite{kim2019cmos,xue2022quantum}, reconfigurable qubit connectivity~\cite{gyger2021reconfigurable}, and heralded spin entanglement~\cite{lomonte2021single}. 

This architecture can be readily extended to other promising solid-state quantum memory platforms. Particularly suitable for heterogeneous integration would be the thin film Si waveguide and a cavity containing isolated color centers~\cite{prabhu2022individually,saggio2023cavity,higginbottom2022optical}; focused-ion-beam-fabricated yttrium orthovanadate crystal comprising rare-earth ion~\cite{ruskuc2022nuclear}; thin film SiC membrane with 4H-C color centers~\cite{crook2020purcell,lukin20204h}; semiconductor quantum dot in a film~\cite{barik2018topological}; and other emerging material platforms~\cite{kim2020hybrid, elshaari2020hybrid}.

\backmatter
\renewcommand{\thefigure}{S\arabic{figure}}
\setcounter{figure}{0}
\renewcommand{\bibfont}{\scriptsize}
\bibliography{sn-bibliography}
\clearpage

\section*{Methods}\label{Method}
\textbf{Color center creation.} To prepare the diamond substrate (with surface orientation [001], we etched the first 10~$\upmu$m of the electronic grade single crystal diamond plate from Element 6 using plasma etching in Ar/Cl$_{2}$ followed by the O$_{2}$ etching to relieve the strained surface. We performed a plane ion implantation of $^{120}$Sn in Innovion at an effective areal dose of 5$\times 10^{11}$ ions per cm$^{2}$ at 350~keV. The implantation target depth is 86~nm, with a longitudinal straggle of 11.7 nm and a lateral straggle of 10.6 nm, estimated from the simulations of stopping and range of ions in matter (SRIM) simulations~\cite{ziegler2010srim}. After implantation, the diamond was annealed at 1200 $^{\circ}$C in an ultrahigh vacuum furnace $10^{-7}$ mbar, followed by a boiling mixture of 1:1:1 sulfuric acid, nitric acid, and perchloric acid cleaning process to remove any graphitic layer due to high temperature annealing. The implantation was at a high dose to guarantee more than one quantum emitter in each quantum channel on average. We estimated a SnV- creation efficiency of approximately 12±5\% in this process. This estimation is based on the spectrum characterization of the diamond pillar array following diamond fabrication, details of which will be elaborated in the subsequent paragraph.

\textbf{Diamond QMC fabrication.} After the generation of SnV$^-$ in the diamond, we deposited 180~nm of silicon nitride (Si$_{3}$N$_{4}$) hard mask using plasma-enhanced chemical vapor deposition, which was patterned using a ZEP-520A electron beam resist with ESpacer conductive polymer and CF$_{4}$ reactive-ion etching. We etched the Si$_{3}$N$_{4}$ pattern in the diamond layer by inductively coupled oxygen reactive ion etching (RIE), followed by an 18~nm of conformal alumina atomic layer deposition. After a breakthrough of alumina with CF$_{4}$ RIE etching, we used a zero-bias oxygen plasma to isotropically undercut the diamond QMC. The silicon nitride mask, as well as the alumina mask, were removed in hydrofluoric acid~\cite{mouradian2017rectangular}.

\textbf{QMC spin-photon interface design.} A perturbative cavity design is implemented in each quantum channel as a dielectric antenna optimized for free space collection. These diamond cavities were connected in groups of 16 via mechanical trusses to parallelize the subsequent integration process. The perturbative cavity is simulated and optimized with finite-difference time-domain (FDTD) simulation in Lumerical. The collection efficiency, a function of the NA = $\sin(\theta)$ of the collection optics, is defined as $\eta(\text{NA})=\frac{\int_0^{2\pi}\int_0^\theta \abs{\vec{E}(\theta,\phi)}^2\text{d}\theta \text{d}\phi}{\int_0^{2\pi}\int_0^{\pi/2}\abs{\vec{E}(\theta,\phi)}^2\text{d}\theta \text{d}\phi}$, where $\vec{E}(\theta,\phi)$ is the far-field of the antenna, as shown in the inset of the supplementary information Fig.~\ref{figS6}a. The Purcell factor of the nanostructure is defined as F$_p=\frac{3}{4\pi^2} (\frac{\lambda}{n})^3(\frac{\text{Q}}{\text{V}})$, where $\lambda/n$ is the wavelength within the refractive index cavity material $n$, and Q and V are the quality factor and mode volume of the cavity respectively. The mode volume of the cavity mode mode is defined as V$=\frac{\int\varepsilon E^2\text{d}V}{\max(\varepsilon E^2)}$, where $\varepsilon$ is the dielectric constant of the material and $E$ is the electric field of the cavity mode. The cavity F$_p$ can also be calculated with F$_p=(\frac{\tau_\text{bulk}}{\tau_\text{on}}-\frac{\tau_\text{bulk}}{\tau_\text{off}})/\xi_\text{ZPL}$. The $\xi_\text{ZPL}$ is about 0.36, representing the fraction of the total emission in the strongest of the two ZPL transitions visible at the temperature of 4~K for SnV$^-$ in the bulk diamond~\cite{kuruma2021coupling}. The implanted SnV dose density yields an average of 3 SnV$^-$ per resonant dielectric antenna within the mode volume.


\textbf{CMOS Post-processing.} We post-processed the CMOS chip with multiple photolithography steps. We did the first photolithography on the chip to define the chiplet locking structure using the alignment markers on the CMOS chip. We removed the fluorescent surface layer on the CMOS chip by CF$_{4}$ reactive-ion etching (RIE) and performed photolithography followed by RIE etching of the spacer material to form the QMC locking structure. We applied a combination of iron chloride, ammonium hydroxide, and hydrogen peroxide to wet etch the top metal routing layer, barrier layer, and via, respectively, forming the QMC locking structure made of oxide in the CMOS chip. We then performed a second photolithography and RIE etching, forming the QMC platform region for the QMC placement. We did the third lithography for the subsequent etching to expose the wire-bonding region. After QMC transfer, we performed additional photolithography followed by e-beam evaporation of a 50~nm Au/5~nm Ti metal layer. This metal layer not only provided a better bonding interface on the CMOS copper, but also allowed for the capacitive strain tuning of the quantum emitters~\cite{wan2020large, clark2023nanoelectromechanical}. 

\textbf{Emitter number per quantum channel calculation.} The n$_\text{emitter}$ is the average number of resonant quantum emitter per frequency channel within a quantum channel. To estimate the total number of quantum emitters across the 11 frequency channels, we sum up all spatially resolvable spots resonating with each channel while tuning emitter frequency. We identified $N_\text{spot}$ = 52322 spots, which detailed data are accessible in the Github repository in the Architecture modeling session at the end of the Methods. Taking into account that each SnV- is counted twice due to its two spin transition tuning across the neighbor frequency channels on average, the total number of distinct spots ($N_\text{spot}^{'}$) is approximately $N_\text{spot}$/2 = 26161. Dividing this by the 11 frequency channels ($k_\text{max}$), we get an average of $N_\text{spot}^{'}$/$k_\text{max}$  = 2378 qubits per frequency channel. Further, dividing this by the total number of nanobeams ($N_\text{sys}$ = 1024) gives an average of about n$_\text{emitter}$ = $N_\text{spot}^{'}$/($k_\text{max}\times N_\text{sys}$) = 2.3 resonant quantum emitters per quantum channel per nanobeam. To estimate the size of the connected qubit graph, we employ the formula: $N_\text{qubit} = n_\text{emitter} \times N_\text{sys} \times p_c \times k_\text{max}$. Here, $p_c$ represents the proportion of the fully connected graph relative to the total number of nodes, a relationship graphically illustrated in Figure 5c of the manuscript.


\textbf{Scalable heterogeneous integration setup and process} The parent diamond was placed on a cut polydimethylsiloxane (PDMS) gel attached to a glass slide as shown in Appendix A. The CMOS chip was mounted on a motorized stage with controllable horizontal ($x$) and lateral movement ($y$) as well as rotation ($\psi$). The diamond-attached glass was screwed to a 3-axis motorized stage with 3-axis piezo control along with the pitch and roll angle control knobs to align the surface of the diamond parallel to the CMOS chip surface. The chiplet alignment can be viewed through a microscope with variable magnification and a long working distance objective. As depicted in Fig.~\ref{fig2}b, we aligned the parent diamond toward the CMOS chip surface using motorized control and moved down the diamond, allowing the QMC array to lock to the post-processed CMOS surface microstructure. We then horizontally moved the parent diamond using the piezo control to release the QMC array. The fabricated QMC locking structure restricted relative movement between the QMC and the parent diamond, breaking the weak hinge (about 150~nm width) between the QMC and the parent diamond in parallel. This method accelerated the heterogeneous integration process by avoiding breaking and picking an individual chiplet. Each QMC can be finely adjusted with probes for better alignment with the bottom metal, and the QMC's design can vary in different regions.


\textbf{Experiment setup} Our cryogenic system employs a closed-cycle helium cryostat with a base temperature of 4K (Montana Instrument), along with a custom cryogenic microscope objective (100$\times$, a numerical aperture (NA) of 0.9) for free space collection. Within the cryogenic system, we use three-axis nano position steppers (Attocube ANP-x,z-50) to move the sample, which is glued to a custom cryogenic printed circuit board (PCB). A programmable voltage source (Keithley 2400) is utilized to apply the voltage to the PCB. The collected signal can be directed to a fiber, a free-space single-photon counting module, or the duo camera system, comprised of a PBS connected to both a scientific CMOS camera (Thorlabs CS235U) and an EMCCD (Hamamatsu ImagEM X2) with an LP filter.

\textbf{Lifetime measurement.} The lifetime measurements were performed using a pulse pattern generator (Anritsu MP1763B), where the APD signal (SPCM-AQRH-16) and the trigger signal of the pulse pattern generator were connected to the PicoHarp 300 for data analysis. Confocal collection fluorescence in free space is filtered using an LP filter (Thorlabs FEL0650). In the lifetime measurement, the laser was modulated by an amplitude modulator (EOSpace, 20GHz bandwidth Lithium Niobate Electro-Optic Modulators) with a digital Mach-Zehnder (MZ) modulator bias controller (iXblue). We biased the MZ modulator to allow a minimum light output and connected the RF control port of the amplitude modulator to the pulse pattern generator. The pulse pattern generator could generate a 2V pulse that turned on the laser within 500~ps providing pulse laser excitation. The pulse trigger signal was analyzed in conjunction with the APD signal received in the PicoHarp 300 to determine the relation between the count rate and the delay time.

\textbf{Laser pulse control.} The tunable resonant laser (MSquared SolsTis with an external mixing module) is modulated with an acoustic-optic modulator (AOM) for pulsing and an electrical-optic modulator (EOM) for frequency sideband tuning. A 515~nm cobalt laser is used to off-resonantly repump the SnV$^-$. The full cycle of the sequence in the main text Fig.~\ref{fig4}b is a sample measurement. We employ two different signal generators (Rohde \& Schwarz SMIQ03 and SMIQ06) with a fast RF switch (Mini-Circuits ZASWA-2-50DR+) to control the modulation signal fed to the EOM. The signal from the signal generator is amplified using a high-power amplifier (Mini-Circuits ZHL-16W-43-S +) operating in the frequency range 1.8~GHz to 4~GHz, which is within the EOM PLE range shown in Fig.~\ref{fig4}b. Controlling the RF switch allows us to switch between the R$_\downarrow$ and R$_\uparrow$ channels, with individual RF amplitudes controlled by two different signal generators. The AOM can simultaneously turn off the R$_\downarrow$ and R$_\uparrow$ channels. In each sample measurement, we have three APD read-out time-bins. APD bin 1 is positioned when the laser is in R$_\downarrow$ channel while during APD bin 2 the laser is also the same. The delay between APD bins 1 and 2 is 50~$\upmu$s. APD bin 3 has another delay time 50~$\upmu$s of from APD bin 2, during which the laser is also in the R$_\uparrow$ channel. 


\textbf{State preparation and measurement error.} 
The preparation and measurement error e$_\text{spam}$ is calculated based on the histogram result P$_\text{dark}(\text{N})$ and P$_\text{bright}(\text{N})$, where we choose an optimum measurement threshold N$_m$ to minimize  e$_\text{spam}=\min(0.5\times(\Sigma_{\text{N}=\text{N}_m}^{\max(\text{N})}\text{P}_\text{dark}(\text{N})+\Sigma_{\text{N}=0}^{\text{N}_m-1}\text{P}_\text{bright}(\text{N})))$. The e$_\text{spam}$ account for errors due to the bright state reading as dark or the dark state reading as bright during preparation and measurement. The average value of P$_\text{dark}(\text{N})$ is 1.6 while the average value of P$_\text{bright}(\text{N})$ is 18. We fit the histogram with the Poisson distribution as shown in Fig.~\ref{fig4}b. This measurement can also be performed when the opposite spin state of SnV$^-$ is prepared.


\textbf{Coherent photon detection efficiency p$_\text{det}$ calculation.} To estimate the detection efficiency, we consider the SnV$^-$ in bulk (without any cavity enhancement). The quantum efficiency of SnV$^-$ emitting photons is 80\%, and four ZPL lines represent 57\% of the energy of the emitted photon. The spin-conserved C line is the bright line, which we are particularly interested in, representing 80\% of the emission energy among all ZPL lines~\cite{aghaeimeibodi2021nanophotonic}. The PSB of the emission is 43\%, and the PSB after applying a 637~nm long-pass (LP) filter is 35\%. Consequently, the intensity ratio between the ZPL C line emission and PSB after the 637~nm LP filter is approximately 1.3. In our previous measurement in Fig.~\ref{fig4}c, we observed an average of 18 counts in PSB after 633~nm LP filter within 50~$\upmu$s readout time. The SnV$^-$’s lifetime is about 5~ns so it can emit a maximum of $\text{photon}_{\text{total}}$ = 10,000 photons in the read-out time. We estimate that we can detect ZPL C line photons $\text{photon}_{\text{ZPL}}$=24 with our 65\% quantum efficiency APD during this cycle. Therefore, we estimate our p$_\text{det}=\text{photon}_{\text{ZPL}}$/photon$_{\text{total}}$ to be greater than 2.4$\times$10$^{-3}$. With higher efficiency detection methods, such as SNSPD~\cite{zhu2016superconducting}, we can achieve a p$_\text{det}$ greater than 3.5$\times$10$^{-3}$.

\textbf{Fully-connected qubit number ratio versus system tunability} To estimate the fully connected qubit number ratio p$_\text{c}$ in Fig.~\ref{fig5}, we perform the Monte Carlo simulation with uniform distributed N$_\text{qubit}$ random emitter zero bias frequency in the inhomogeneous range $v_\text{inh}$ from 0\% to 100\%. The maximum tunability for each individual emitter is assumed to be a zero-mean Gaussian distribution, and the standard deviation is fitted to match the experimental sampled mean value for the absolute tuning range. And with a dynamic programming calculation counting the number of the quantum emitter that can be connected within all the possible emitters' tuning ranges, we can estimate the number of connected node N$_\text{con}$ among N$_\text{qubit}$ samples we have. This value is an average of 10 times in the Monte Carlo simulation. The tunability ratio is also sweep to provide the Fig.~\ref{fig5}b curve of p$_\text{c}$ against $\overline{\Delta v_\text{m}}$/$ v_\text{inh}$ under various N$_\text{qubit}$. Although we use 11 frequency channels to cover a 20~GHz emitter spectral range, $v_\text{inh}$ can reach a 150~GHz for SnV$^-$, which makes our $\overline{\Delta v_\text{m}}$/$ v_\text{inh}$ to 1.3\%. When expanding the spectral range of the emitter, we also expect to see much more numbers for N$_\text{qubit}$ in the same field of view.

\textbf{Architecture modeling.} We provide a hardware architecture modeling file, accessible at \url{https://github.com/LinsenLi97/Freely-scalable-architecture-modeling}. Included in this repository is a Python notebook covering various topics: 1) Key parameters for general-purpose quantum computing; 2) Architectural entanglement sequence for estimating entanglement rates; 3) Spectral diffusion's impact on entanglement rate; 4) Effect of spectral diffusion and Purcell enhancement on visibility; 5) Strain tuning models for ZPL shift and spin splitting; 6) Thermal budget considerations for the architecture's operation; 7) Emitter statistical data. This modeling resource helps to understand the practical compromises among various architectural parameter settings.


\textbf{Acknowledgments.} This work was supported by the MITRE Corporation Quantum Moonshot Program, the NSF STC Center for Integrated Quantum Materials (DMR-1231319), the ARO MURI W911NF2110325 and the NSF Engineering Research Center for Quantum Networks (EEC-1941583). L.L. acknowledges funding from NSF QISE-NET Award (DMR-1747426). L.D. and C.E-H. acknowledge funding from the European Union’s Horizon 2020 research and innovation program under the Marie Sklodowska-Curie grant agreements No.840393 and 896401. I.H. acknowledges the funding support by the National Science Foundation RAISE-TAQS (Grant No. 1839159). K.C.C. acknowledges funding support by the National Science Foundation Graduate Fellowship. M.E.T acknowledges support from the ARL ENIAC Distinguished Postdoctoral Fellowship. The authors thank Adrian Menssen for the discussion of the SLM application. The authors thank Yong Hu for the comments on the figure. The authors thank Ruicong Chen for help with the foundry tape out and Eric Bersin for useful comments on the manuscript. L.L. thanks Shiyin Wang for the kind support for him.

\textbf{Author contributions.} L.L. conducted device design and fabrication, scalable heterogeneous integration, and measurements of the whole experiment. L.D.S. contributed to the cryogenic measurement setup and the experiment automation code. I.H. performed the tape out of the CMOS chip and contributed to the room-temperature measurement setup. K.C.C. contributed to the diamond fabrication and cryogenic measurement. Y.S., I.C., M.T., Y.H., and Y.G. contributed to the setup of the cryogenic experiment. H.C. contributed to the theoretical analysis of the architecture, C.E.H. contributed to the transfer print stage for scalable transfer, J.D. contributed to the transfer and the scientific plotting, and Y.G. performed the COMSOL simulation for the strain tuning. G.C. contributed to characterizing the linewidth of the color center to verify the color center annealing properties. M.I.I and R.H. provided an experience for the CMOS tape out. D.E. conceived the idea and supervised the whole project. D.E. and L.L. designed the experiment, and L.L. wrote the manuscript with input from all authors.

\textbf{Competing interests.} The authors declare no competing interests.

\pagestyle{plain}
\onecolumn 
\appendix
\section{Large-scale heterogeneous integration}\label{secA}

\begin{figure*}[b]%
\centering
\includegraphics[width=1\textwidth]{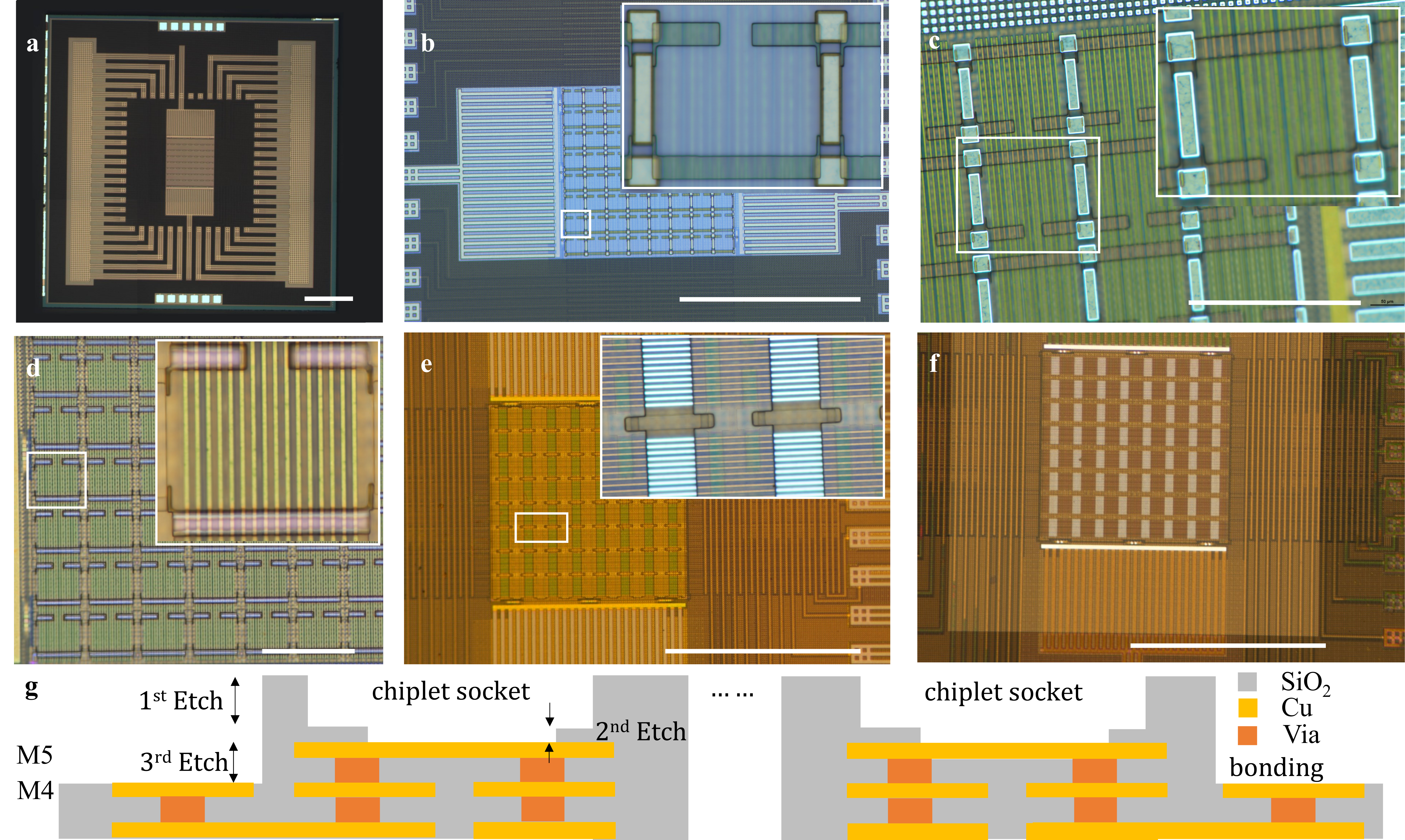}
\caption{\textbf{CMOS post-processing procedure.} \textbf{a,} Optical microscope image of the bare die CMOS chip from the foundry (scale bar: 500$\upmu$m). \textbf{b,} Optical microscope image after the first photolithography (scale bar: 500$\upmu$m). \textbf{c,} Optical microscope image after the first dry etching (scale bar: 100$\upmu$m). \textbf{d,} Optical microscope image after wet etching (scale bar: 100$\upmu$m). \textbf{e,} Optical microscope image after second photolithography and dry etching (scale bar: 500$\upmu$m). \textbf{f,} Optical microscope image after third photolithography (scale bar: 500$\upmu$m). \textbf{g,} Cross-section of the post-processed CMOS chip.}\label{figS1}
\end{figure*}

Figure~\ref{figS1}a presents the optical image of the 3~mm $\times$ 3~mm bare die of the CMOS chip. Alignment markers on the chip were used for the first photolithography that created the initial photoresist layer defining the locking structure of the chiplet, as seen in Fig.~\ref{figS1}b. The passivation layer on the CMOS chip was removed using CF$_{4}$ reactive ion etching (RIE) and was followed by photolithography and RIE etching of the spacer oxide to form the QMC locking structure, illustrated in Fig.~\ref{figS1}c. The key feature for the QMC locking structure is slightly larger than the QMC's edge, so it can serve as the nanostructure that restricts the QMC movement during the lock and release. Wet etching was performed with a solution of iron chloride, ammonium hydroxide, and hydrogen peroxide to remove the top routing metal layer, barrier layer, and via, respectively, resulting in the QMC oxide locking structure displayed in Fig.~\ref{figS1}d. The second photolithography and RIE etching formed the QMC platform region for QMC placement, as shown in Fig.\ref{figS1}e and its inset. The third lithography, depicted in Fig.~\ref{figS1}f, prepared the chip for subsequent etching to expose the internal routing metal layer for the wire bonding region. Finally, Fig.~\ref{figS1} g shows the cross-section of the post-processed CMOS chip, with M3-M5 corresponding to metal layers 3 to 5 in the foundry metal definition; this CMOS process has a total of 6 metal layers.

\begin{figure*}[ht]%
\centering
\includegraphics[width=1\textwidth]{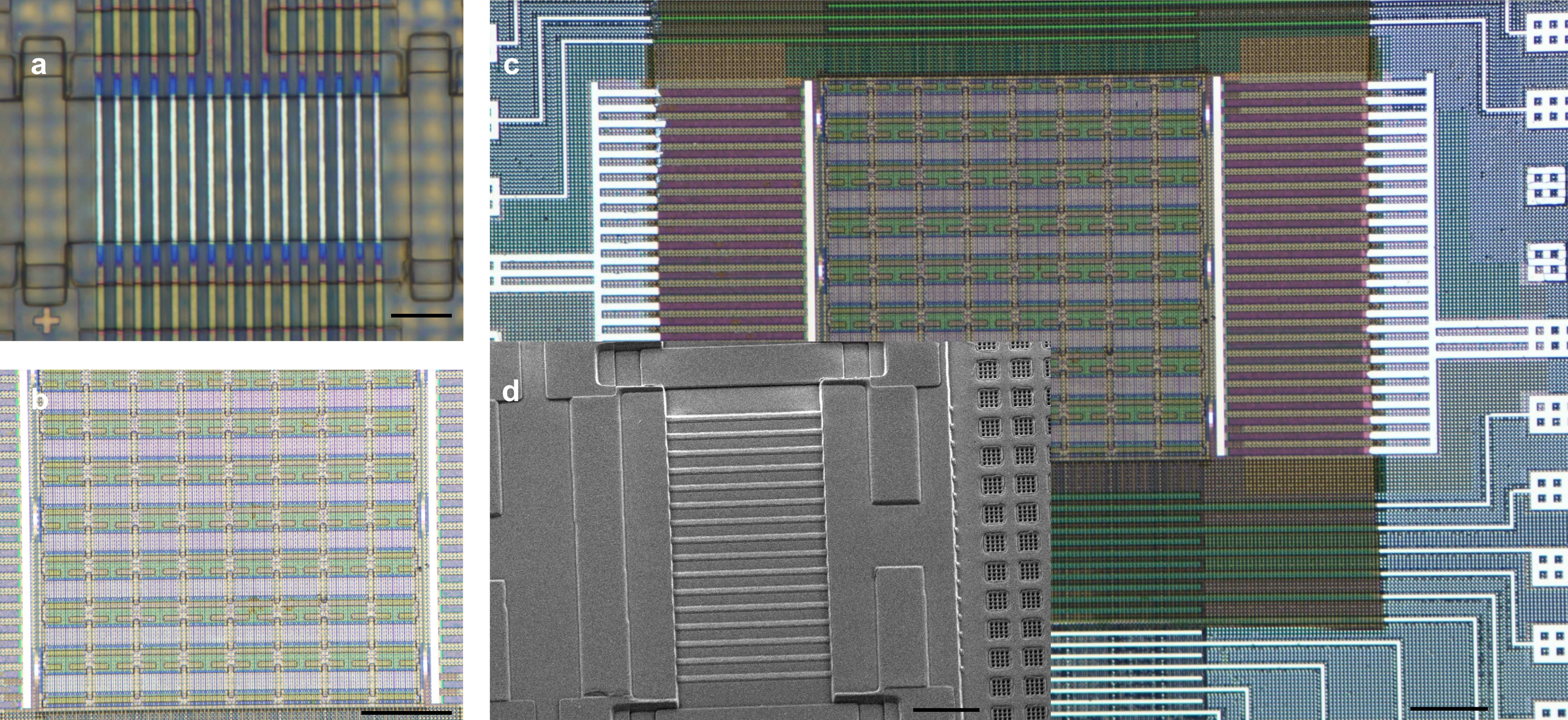}
\caption{\textbf{CMOS post-processing result.} \textbf{a,} Optical microscope image of a single post-processed CMOS socket unit (scale bar: 10$\upmu$m). \textbf{b,} Optical microscope image of an 8~$\times$~8 array of post-processed CMOS sockets (scale bar: 100$\upmu$m). \textbf{c,} Optical microscope image of the central region of the post-processed CMOS chip (scale bar: 100$\upmu$m). \textbf{d,} SEM image of the post-processed CMOS socket unit (scale bar: 100$\upmu$m).}\label{figS2}
\end{figure*}

\begin{figure*}[ht]%
\centering
\includegraphics[width=1\textwidth]{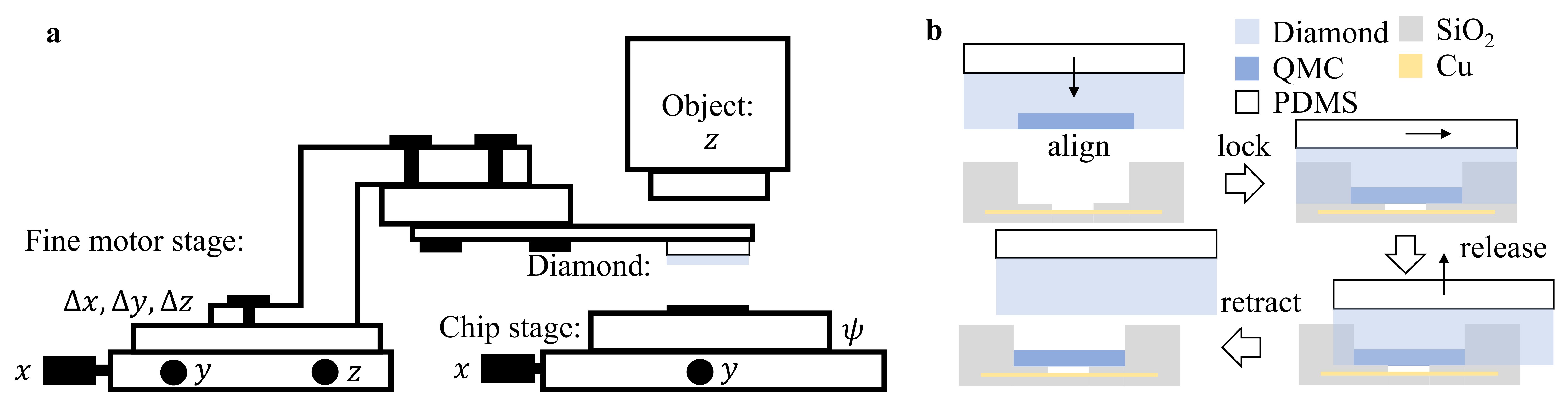}
\caption{\textbf{Detailed of lock-and-release transfer operation for QSoC fabrication.} \textbf{a,} Setup for the scalable heterogeneous integration process between the fabricated parent diamond substrate and the target CMOS chip. \textbf{b,} Cross-section of the unit QMC and its corresponding CMOS locking structure during the transfer process. For the GIF view, see \url{https://github.com/LinsenLi97/Freely-scalable-architecture-modeling/blob/main/lock-and-release-animation.gif} for the cross section animation and \url{https://github.com/LinsenLi97/Freely-scalable-architecture-modeling/blob/main/lock-and-release-3D.gif} for the 3D animation.}\label{figS3}
\end{figure*}

 Figure~\ref{figS2} displays the surface of the post-processed CMOS chip, featuring the QMC locking structure on top of the chip, which facilitates the scalable transfer process. Figure~\ref{figS2}a illustrates a single CMOS socket unit, while Fig.~\ref{figS2}b presents an 8~$\times$~8 array of CMOS sockets within the central 500~$\upmu$m $\times$ 500~$\upmu$m area of the chip. A larger view of the chip is provided in Figure~\ref{figS2}c, and Fig.~\ref{figS2}d offers the SEM of the surface of the post-processed CMOS chip. Figure~\ref{figS3} shows the scalable transfer setup, which allows for scalable heterogeneous integration. This figure also shows the setup and the changes in the cross-section of the scalable heterogeneous integration process.
 
\section{Programmable spin-photon interface}\label{secB}

In addition to the Purcell factor and the far-field collection efficiency discussed in the Methods section, another crucial parameter is the fraction of emitted power propagating in the $+z$ direction, defined as the T$_z=\frac{\text{P}_z}{2\text{P}_x+2\text{P}_y+\text{P}_z}$. P$_x$ and P$_y$ represent the transmitted power in the $x$ and $y$ directions, respectively, while P$_z$ denotes the transmitted power in the $z$ direction, assuming a perfect metal reflector at a distance of $\Delta z$ below the diamond cavity in the FDTD simulation. A dipole source oriented in the $y$ direction (perpendicular to the nanobeam) is used in the center of the cavity region to excite the cavity mode. The final cavity design features a diamond waveguide with a width (W) of 280~nm and thickness (H) of 200~nm, periodically patterned with holes of 64~nm radius and a 202~nm spacing (a). The defect supporting the cavity mode is introduced by the Gaussian modulation x(m)=d$_0$+a$\times$[1-exp(-$\frac{\text{m}^2}{2\sigma^2})]$, where x(m) denotes the distance of the m$^{th}$ holes on one side from the center of the cavity. d$_0$ is the distance from the center of the first hole to the center of the cavity, and $\sigma$ is a dimensionless number controlling the distribution of the location of the holes. The optimized parameters are d$_0=$ 113~nm and $\sigma=$~4.6. To improve the collection efficiency in free space, half-round perturbation bumps are introduced along the cavity region, which acts as a grating. Four pairs of half-round bumps are placed on each side of the cavity. After optimization, the centers of the four bumps are 200~nm, 600~nm, 1070~nm, and 1470~nm from the cavity center, and their radii are 50~nm, 50~nm, 50~nm, and 80~nm, respectively. With these parameters, the resonant dielectric antenna has a quality factor of 2000, a T$_z$ of 99.2\%, a $\eta(0.5)$ of 78\%, and a $\eta(0.9)$ of 96\%, assuming a perfect back metal reflector at a controlled distance of $\Delta z =$ 250~nm. This efficient spin-photon interface is essential for high-speed entanglement generation and the practicality of distillation protocols~\cite{campbell2008measurement,li2021field}.

\begin{figure*}[ht]%
\centering
\includegraphics[width=0.6\textwidth]{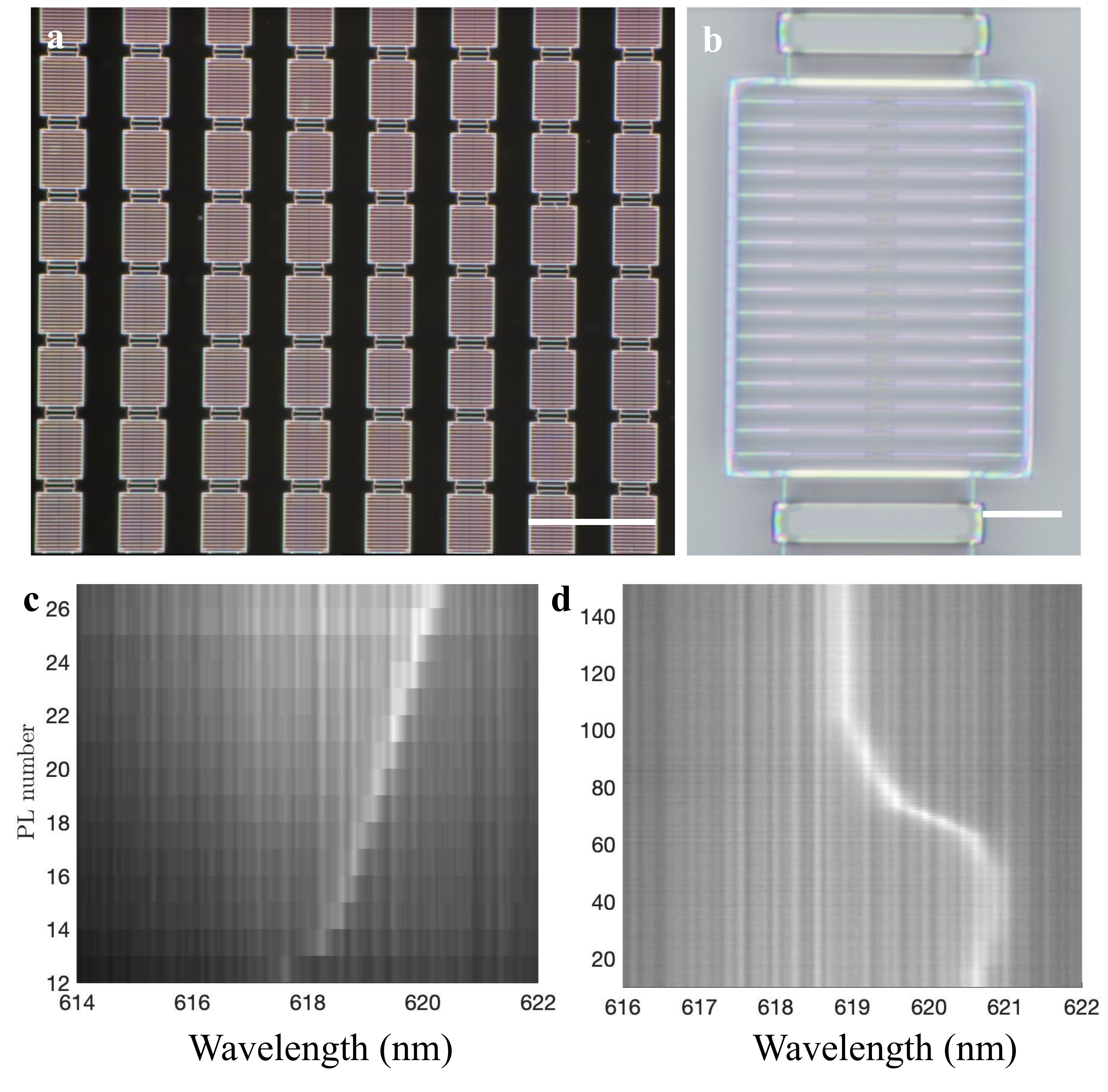}
\caption{\textbf{Diamond QMC array fabrication and Gas tuning process for adjusting the resonant wavelength of the resonant dielectric antenna.} \textbf{a,} Dark-field optical microscope image of the QMC array on the parent diamond (scale bar: 100$\upmu$m). \textbf{b,} Optical microscope image of a single QMC (scale bar: 10$\upmu$m). \textbf{c,} Redshift of the cavity resonance due to gas deposition on the cavity. \textbf{d,} Blue-shift of the cavity resonance resulting from gas desorption using a pulsed high-power laser.
}\label{figS4}
\end{figure*}

Figure~\ref{figS4}a shows the dark-field optical microscope image of the QMC array on the parent diamond, and Fig.~\ref{figS4}b shows the optical microscope image of a QMC with 16 quantum channels. Due to the fabrication variations, the cavity resonant frequency does not exactly match the SnV$^-$ ZPL. The inhomogeneous cavity fabrication variation is overcome by using a gas-tuning method. A supercontinuum laser (SuperK) is used to monitor the cavity spectrum during the gas-tuning process. The gas deposition on the cavity results in a redshift of the cavity resonant frequency, whereas a pulsed high-power laser can achieve the opposite process of gas desorption, causing a blue shift in the cavity resonance. Gas deposition can be applied globally to all the cavities, and the laser-assisted desorption process can be performed individually for each cavity. The gas-tuned cavity remains stable in a cryogenic environment without significant temperature fluctuations~\cite{bhaskar2020experimental}. Figure~\ref{figS4} also illustrates the gas adjustment process to adjust the resonant wavelength of the perturbative cavity with the supercontinuum laser (SuperK) monitoring the cavity reflection spectrum. Figure~\ref{figS4}c demonstrates the gas deposition process within the cavity, causing a redshift in the resonant frequency of the cavity. A larger PL number indicates a later measurement time during the gas deposition process. Figure~\ref{figS4}d shows the gas desorption process of the cavity using a pulsed high-power laser, which results in a blue shift of the cavity resonance.

\begin{figure*}[ht]%
\centering
\includegraphics[width=1\textwidth]{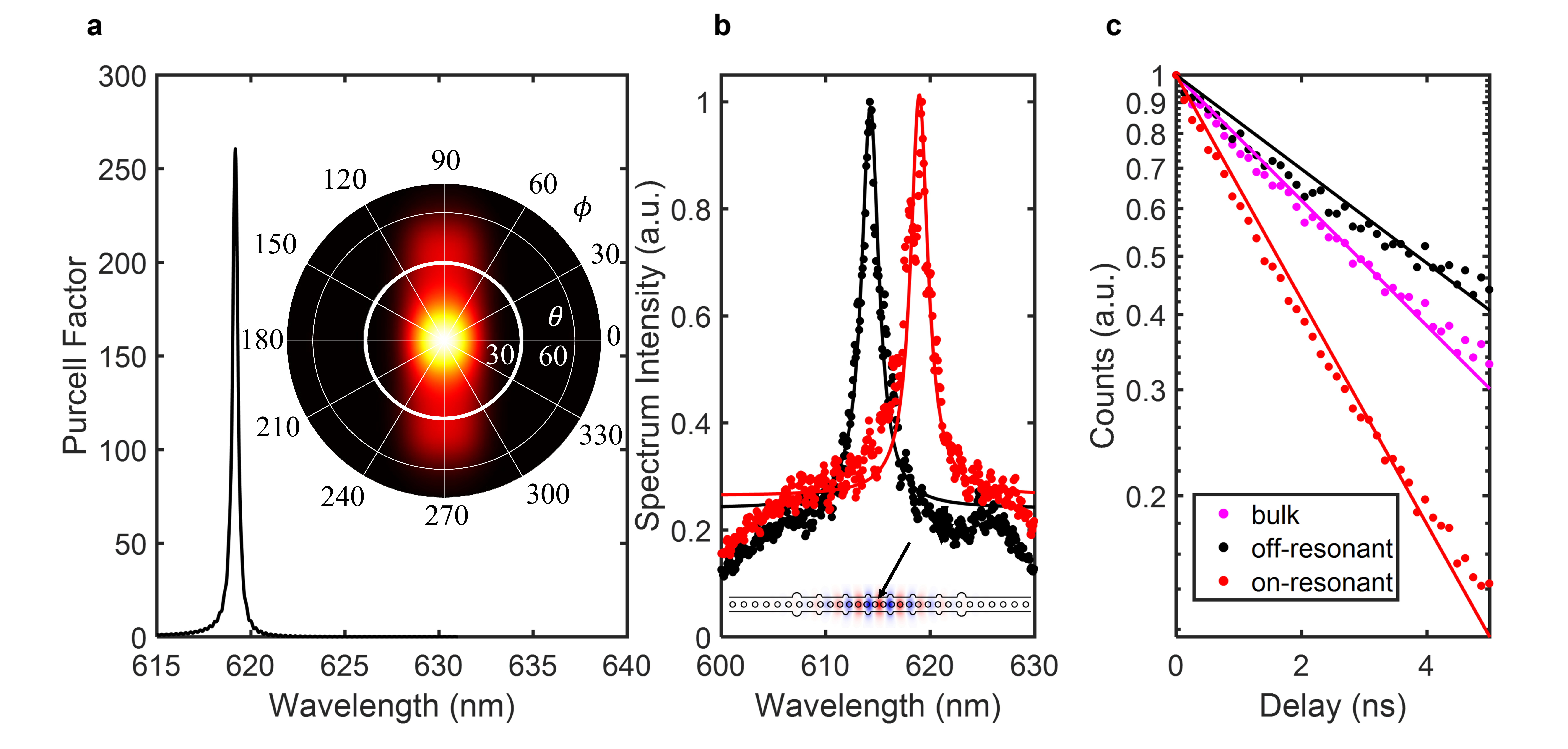}
\caption{\textbf{The Purcell enhancement of the resonant dielectric antenna}. \textbf{a,} Simulated Purcell factor and far-field distribution of the quantum emitter in the center of the optimized resonant dielectric antenna. \textbf{b,} Resonant dielectric antenna spectrum (interference between laser reflection and resonant dielectric antenna reflection) displaying off-resonant (black) and on-resonant (red) peaks with the SnV$^-$ ZPL wavelength. The inset shows the simulated electric field distribution (depicting the real part of the electric field, with red being positive and blue being negative) overlaid with the edge of the resonant dielectric antenna design, assuming that the dipole is at the center of the resonant dielectric antenna with an in-plane orientation perpendicular to the nanobeam. \textbf{c,} Representative lifetime measurements when the resonant dielectric antenna is on/off-resonant with the SnV$^-$ ZPL after/before gas tuning are shown in red/black, respectively. The lifetime of another SnV$^-$ in the bulk diamond is shown in magenta.}\label{figS6}
\end{figure*}

Figure~\ref{figS6}a presents the simulated Purcell factor and the far-field distribution of the electromagnetic field for an emitter optically coupled to the resonant dielectric antenna mode. Our design achieves a collection efficiency of $\eta=$ 96\% for a numerical aperture (NA) of 0.9 and $\eta=$ 78\% for NA = 0.5. As depicted in Fig.~\ref{figS6}b, we tuned the resonant dielectric antenna spectrum through gas deposition~\cite{aghaeimeibodi2021nanophotonic,kuruma2021coupling,evans2018photon} (see Methods and Appendix A). The quality factor of Q = 220 remains unchanged before and after gas deposition. We observe a tuning range $\Delta_\text{gas}>$ 10~nm, and this tuning method can accommodate up to 36~nm without reducing the quality factor~\cite{li2015coherent}. The impact of spectrally tuning the nanophotonic resonance on the lifetime of the emitter is illustrated in Fig.~\ref{figS6}c. We compare the cases of an emitter in the bulk material $\tau_\text{bulk}=4.12$~ns, an emitter in the off-resonance dielectric antenna $\tau_\text{off}=5.56$~ns, and $\tau_\text{on}=2.32$~ns after tuning the dielectric antenna on-resonance with its ZPL.

\begin{figure*}[ht]%
\centering
\includegraphics[width=1\textwidth]{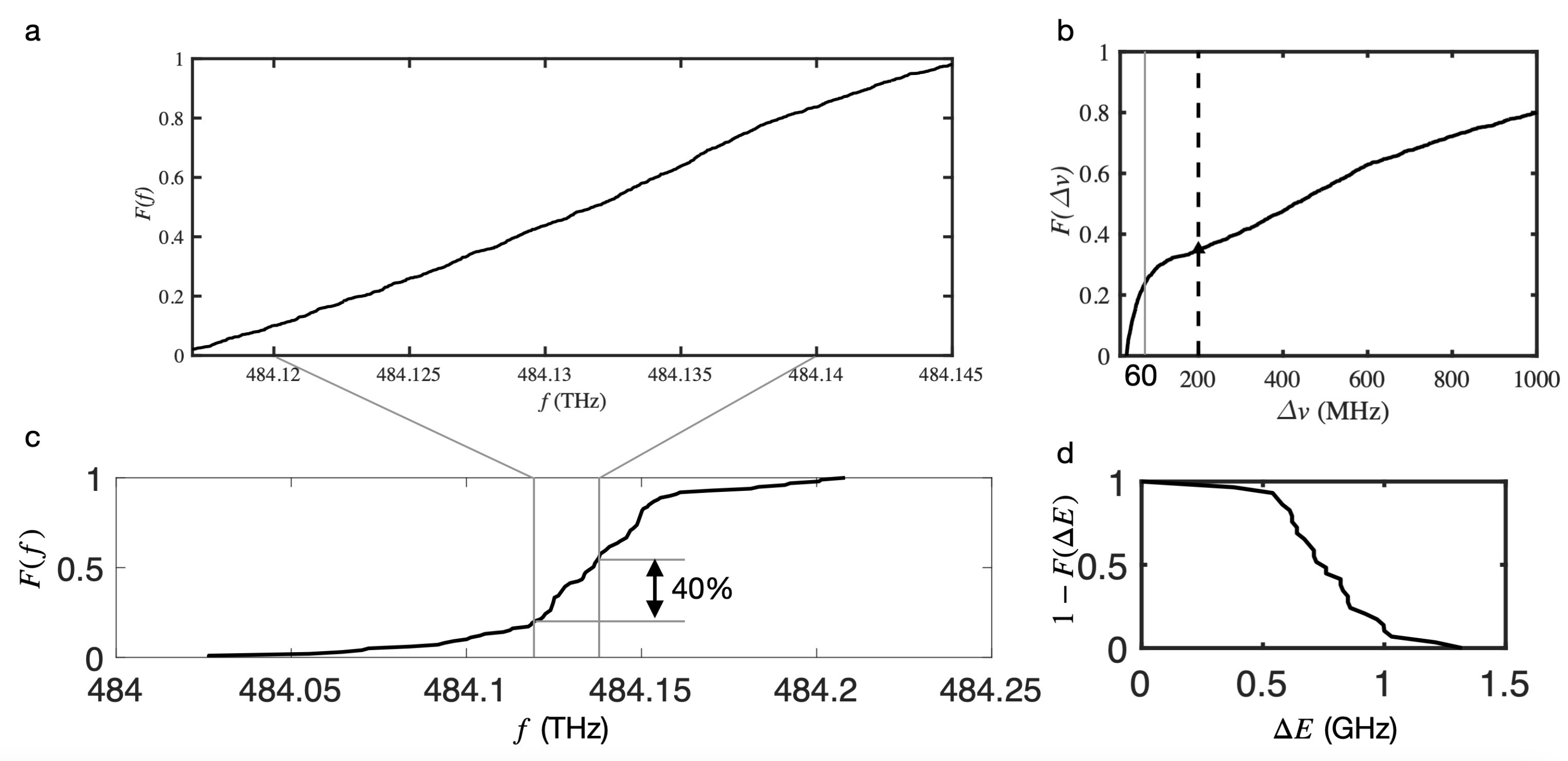}
\caption{\textbf{Diamond quantum emitter statistics}. \textbf{a,} The cumulative distribution function (cdf) of Emitter central frequency f (wide field PLE between 484.115 THz to 484.145THz). \textbf{b,} The cdf of the emitter linewidth smaller than $\Delta$v. \textbf{c,} The cdf of the SnV$^-$ ZPL inhomogenour frequency in a field of view. \textbf{d,} The cdf of the energy splitting larger than $\Delta$E.}\label{figS6new}
\end{figure*}

In our analysis of the statistical widefield PLE data for SnV- centers in the absence of a magnetic field bias, we have illustrated the cumulative distribution function (cdf) of the central frequency f in Fig.~\ref{figS6new}a. This figure shows a nearly linear distribution ranging from 484.115 THz to 484.145 THz for more than 2000 SnVs, suggesting us to postulate a uniform distribution of SnV- centers within this frequency spectrum. Fig.~\ref{figS6new}b displays the cdf of emitter linewidths determined through widefield PLE. As indicated, the measured linewidths exceed the transform-limited linewidth due to spectral diffusion. Still, 20\% of emitters have linewidths (inclusive of spectral diffusion) that are within twice the transform-limited linewidth (30 MHz). In the context of optical addressability for the two spin states, color centers with a frequency variation under 200 MHz are considered suitable, accounting for 35\% of the sample. Fig.~\ref{figS6new}c demonstrates the broad inhomogeneous distribution of SnV ZPL frequencies within a single field of View, aligning with the near-linear ZPL distribution in a narrow range as shown in Fig.~\ref{figS6new}a. Additionally, Fig.~\ref{figS6new}d reveals that 80\% of the widefield PLE splittings under a 0.13 T magnetic field exceed 0.6 GHz, consistent with the data presented in the manuscript.

\begin{figure*}[ht]%
\centering
\includegraphics[width=1\textwidth]{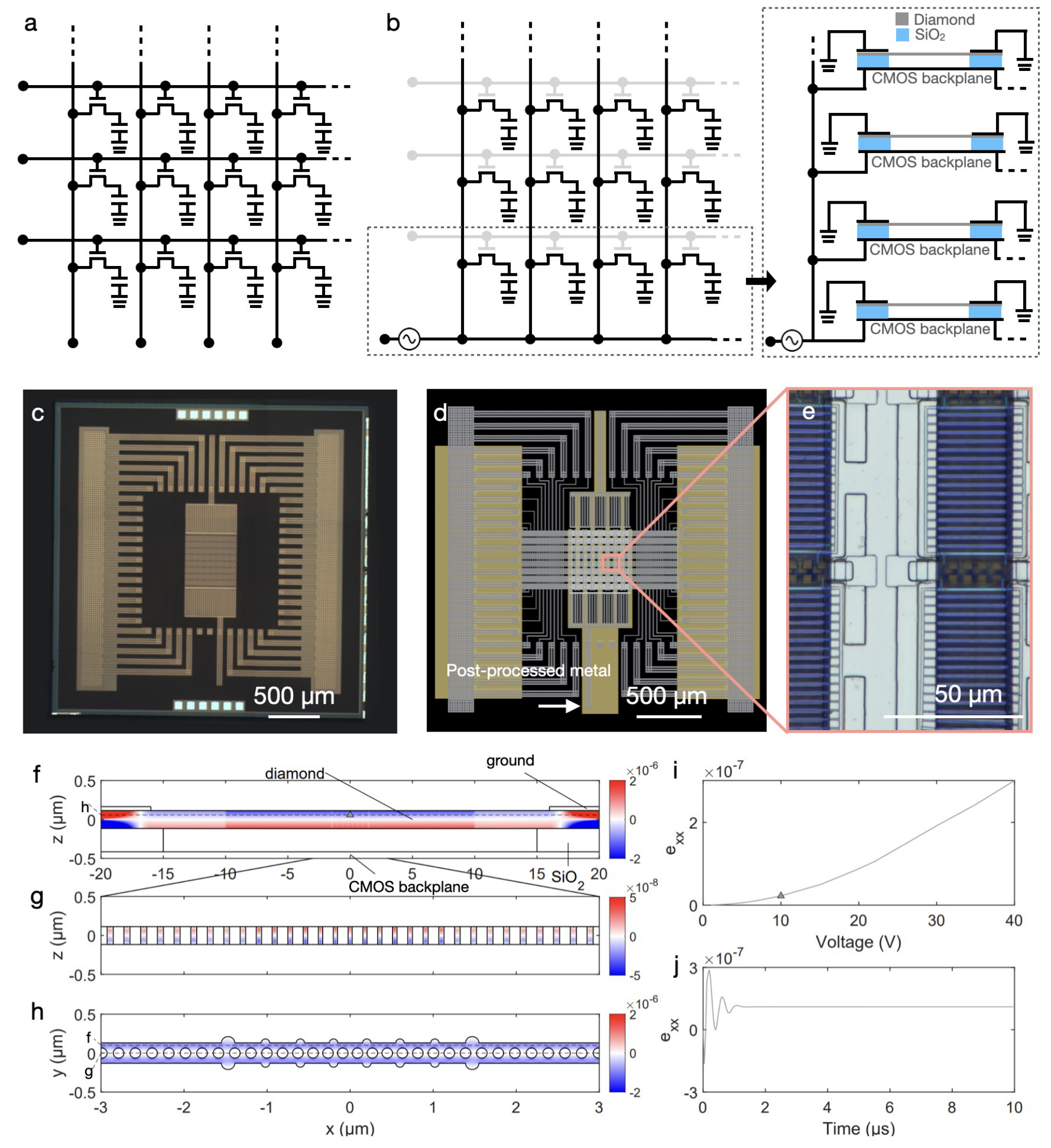}
\caption{\textbf{Strain tuning mechanisim and simulation in COMSOL Multiphysics.} \textbf{a,} The crossbar structure CMOS circuit diagram. \textbf{b,} The circuit in the manuscript, multi-columns shorting together for the global bias control. A cross-section image is shown as the sample of the circuit region on the right. \textbf{c,} The initial CMOS chip from vendor. \textbf{d,} The post-processed CMOS chip design with metalization (labeled in the yellow color, the initial CMOS metal routing is in white color. \textbf{e,} A zoom-in region of the post-processed device layer. \textbf{f,} Device strain tensor XX component e$_\text{XX}$ distribution across the cross-section of the x-z plane at y = 0.1 µm in COMSOL simulation. \textbf{g,} Zoomed-in e$_\text{XX}$ strain distribution in the x-z plane, with the cavity region on the x-axis ranging from  -3 µm to 3 µm and y = 0 µm. \textbf{h,} e$_\text{XX}$ strain distribution in the x-y plane with the cavity region at z = 57.5 nm, illustrated by the dashed line in \textbf{f}. The two dashed lines in \textbf{h} shows the x-z plane position in \textbf{f} and \textbf{g} respectively. \textbf{i,} e$_\text{XX}$ strain relation with bias voltage for the location marked with a gray triangle in \textbf{f}. \textbf{j,} The transient e$_\text{XX}$ response when the bias voltage is 10 V at the location marked with a gray triangle in \textbf{h}.}\label{figS7}
\end{figure*}

Figure~\ref{figS7} presents the result of the simulation of stationary electromechanics conducted using COMSOL Multiphysics. The simulation structure comprises a single diamond waveguide with holes and bumps, as described in the manuscript, an electrode on the top right providing voltage, and the oxide platform region on the CMOS chip to avoid severe deformation and grounding at the base of the substrates. The remaining parts are filled with air and do not contribute to the strain tensor. We apply several constraints and boundary conditions to the stationary solver. In the solid mechanics' portion, the diamond body is in the free deformation state, while the surfaces on both ends, which connect to the CMOS chip in reality, are fixed. We also fix the intersecting surface between the oxide and the ground. For the electrostatics portion, we apply charge conservation to the entire object and add terminal and ground boundary conditions to the electrode and the ground, respectively. In Fig.~\ref{figS7}a, the CMOS circuit's crossbar structure is depicted, capable of applying individual voltage control to the capacitance of each pixel region. Fig.~\ref{figS7}b illustrates the circuit configuration used in our manuscript, where multiple columns are combined to facilitate global chip control. This figure also includes a cross-section of the sample for each quantum channel control on the right. In Fig.~\ref{figS7}c, we present the initial layout of the CMOS chip as received from the vendor, and Fig.~\ref{figS7}d shows the chip post-processing metal location, highlighting the top electrode appearance. Wire bonding is applied to these post-processed metal areas, connecting them to an external Keithley voltage source. Fig.~\ref{figS7}e offers a top view of the device region after post-processing. Fig.~\ref{figS7}f to~\ref{figS7}h demonstrate the simulated strain distribution in the diamond when a bias is applied between the CMOS backplane and the top ground electrode. Specifically, Fig.~\ref{figS7}f shows the device strain tensor XX component e$_\text{XX}$ distribution. We observe two sign-changing nodes along the x-axis, and the strain tensor approaches 0 when z is near 0, and Fig. ~\ref{figS7}g zooms in on this strain distribution. The profile’s edge is slightly curved due to the electric field, and the strain tensor vanishes in the middle along the z-direction. Consequently, we selected a biased plane at z = 57.5 nm marked with the dashed lines in Fig.~\ref{figS7}f, and plot the strain distribution in Fig.~\ref{figS7}h. In Fig.~\ref{figS7}i, we plot the strain curve as a function of voltage. The position of the emitter is denoted by a gray triangle marker in Fig.~\ref{figS7}f. Figure~\ref{figS7}j illustrates the transient strain response observed upon applying a 10 V bias voltage to the electrode. The simulation results suggest the feasibility of strain tuning through the varying voltage applied to the diamond.

\begin{figure*}[ht]%
\centering
\includegraphics[width=1\textwidth]{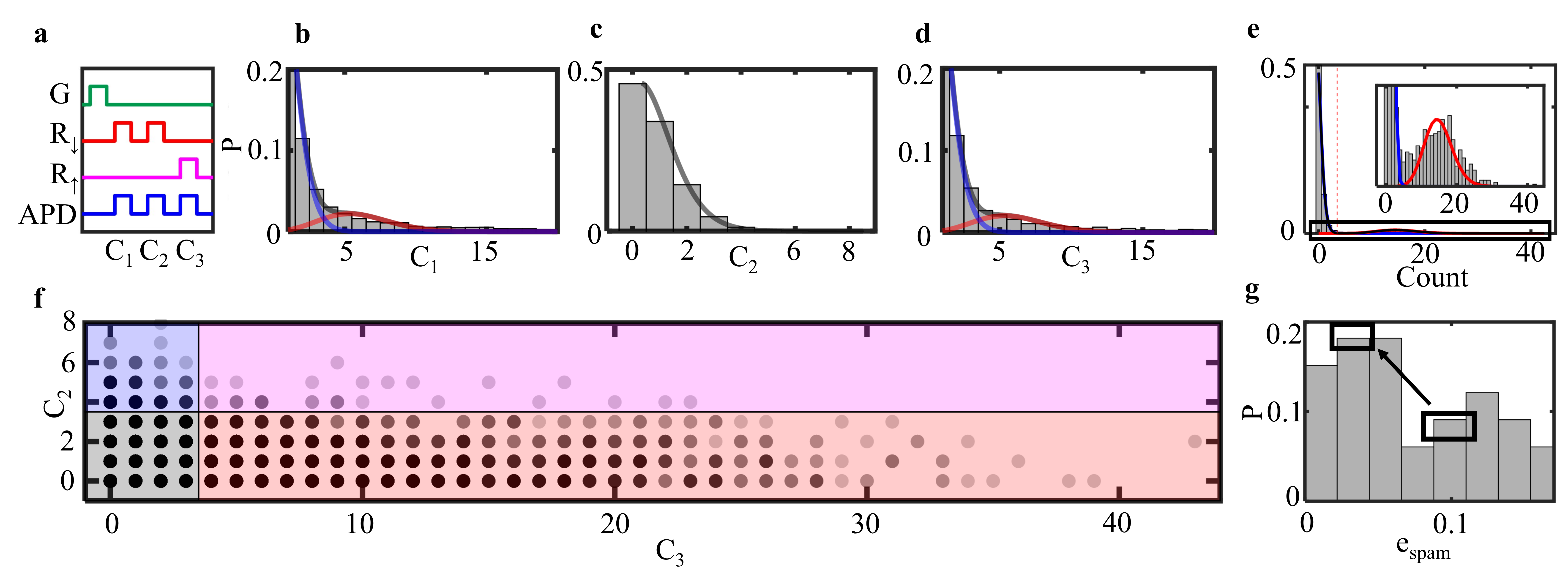}
\caption{\textbf{The spin state measurement result without post-selection}. \textbf{a} Pulse control sequence. \textbf{b-d,} Histogram plots of the spin readout results for the three APD time bins in Fig.~\ref{fig4}b, presented without post-selection. \textbf{e,} Readout histogram without post-selection, featured a zoomed-in histogram in the inset. \textbf{f,} Correlation between the second and third APD time bins, with the correlation plot divided into four quarters, each represented by a different color: gray region (APD 2$<$N$_\text{m}$, APD 3$<$N$_\text{m}$), red region (APD 2$<$N$_\text{m}$, APD 3$>$N$_\text{m}$), blue region (APD 2$>$N$_\text{m}$, APD 3$<$N$_\text{m}$), and magenta region (APD 2$>$N$_\text{m}$, APD 3$>$N$_\text{m}$). N$_\text{m}$ is fit from \textbf{b} here, contrasting with N$_\text{m}$, which is derived from the post-selected data. \textbf{g,} The histogram of the e$_\text{spam}$ statistics without post-selection. The red rectangle represents the change of e$_\text{spam}$ after post-selection corresponding to the data in the manuscript Fig.~\ref{fig4}b.}\label{figS8}
\end{figure*}

Figure~\ref{figS8}a shows the pulse control sequence, while Fig.~\ref{figS8}b-d presents the histogram of the three APD time bin readout results in Fig.~\ref{fig4}a, along with a maximum likelihood estimation of mixture Poisson distribution fitting (black curve), $p(n)=(1-p_0)\frac{\lambda_1^n\exp(-\lambda_1)}{n!}+p_0\frac{\lambda_2^n\exp(-\lambda_2)}{n!}$. Here, $p_0$ is the fitted probability between two distributions, while $\lambda_1$ and $\lambda_2$ represent the average count numbers of the bright- and dark-state events, respectively. The fitted individual Poisson distributions for the bright and dark states are plotted separately in red and blue. We estimate the read-out threshold count N$_\text{m}$  to minimize the read-out error by satisfying $(1-p_0)\lambda_1^{\text{N}_\text{m}}\exp(-\lambda_1)=p_0\lambda_2^{\text{N}_\text{m}}\exp(-\lambda_2)$. Bin 1 and bin 3 are fitted to the mixture Poisson distribution, while bin 2 is fitted to a single Poisson distribution. The fitted N$_\text{m}$ is 3.5 here. Based on the relation between APD count and N$_\text{m}$, the correlation can be divided into four quadrants as shown in Fig.~\ref{figS8}, labeled with the colors gray, red, blue, and magenta. In the gray area, we consider the spin to be not initialized to the correct charge state, since APD bins 2 and 3 are both considered dark in the readout result. In the red area, we consider the successful spin preparation and measurement as expected. In the blue and magenta regions, the readout is considered an error for state preparation and measurement. We repeat the sequence in Fig.~\ref{figS8}a 10000 times and obtain 143 correlated events in the blue region, 1508 correlated events in the red region, and 29 correlated events in the magenta region, indicating a preparation and measurement error of 10.24\% without post-selection of the first APD bin count. The post-selection in the manuscript not only selects the correct spin state for reading out but also chooses the bright spin state when the laser is better aligned with the emitter's resonant frequency. The readout count will be darker if spectral diffusion occurs, causing the emitter's resonant frequency to be misaligned with the laser. This post-selection also raises the readout threshold of N$_\text{m}$ to better distinguish the dark state from the bright state.

\section{Measurement setup and all-to-all routing approach}\label{secC}

\begin{figure*}[ht]%
\centering
\includegraphics[width=1\textwidth]{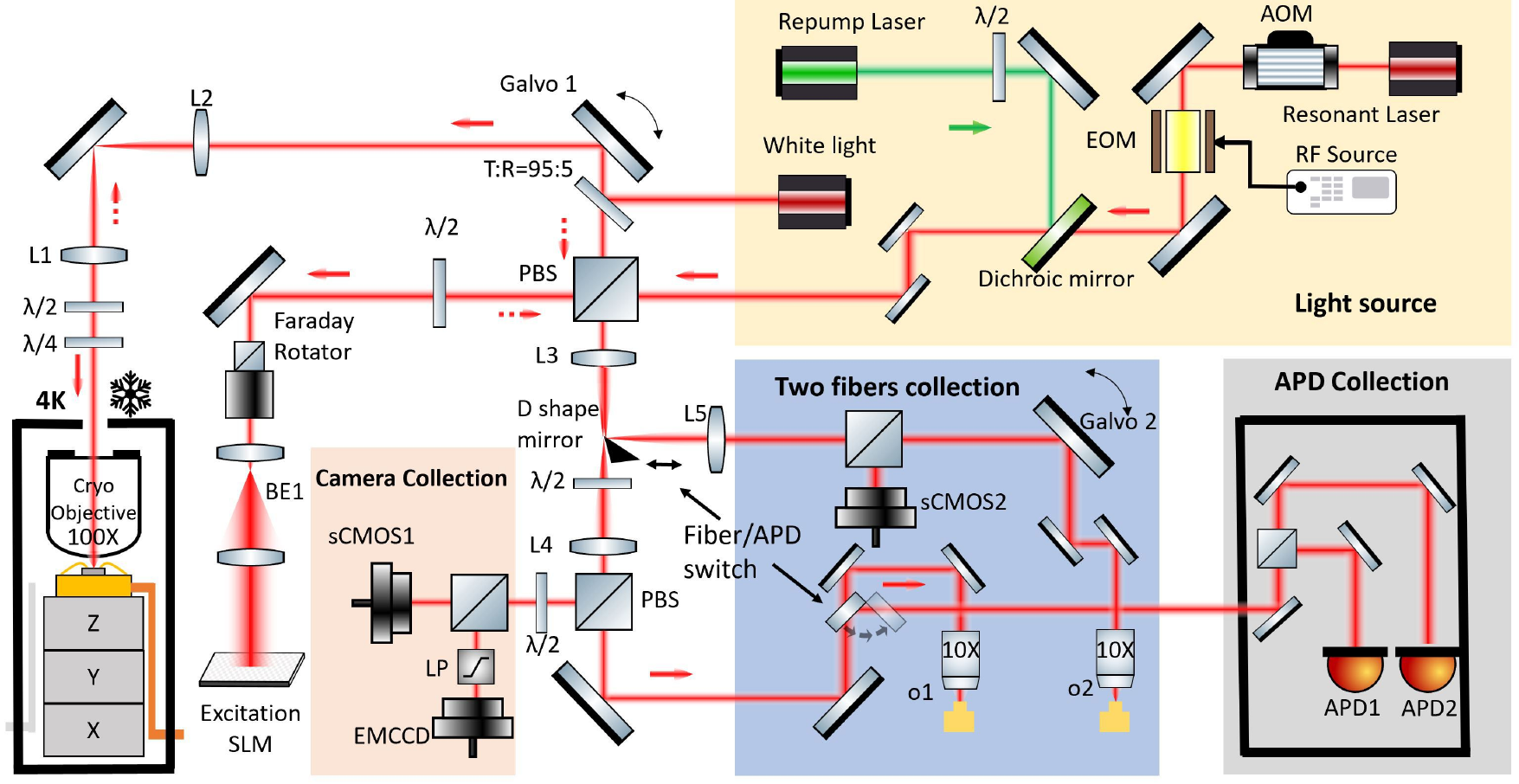}
\caption{\textbf{The optical measurement setup for freely scalable hardware architecture demonstration.}}\label{figS9}
\end{figure*}

Figure~\ref{figS9} illustrates the experimental setup for demonstrating the freely scalable hardware architecture, including the cryogenic system, laser source, 4f confocal system, SLM excitation, collection box including the duo camera image system (EMCCD and scientific CMOS camera), and collection APD detectors. The resonant laser here is modulated by an AOM and EOM, and mixed with the green repump laser via a dichroic mirror. For PLE, the resonant laser is pulsed by an AOM to excite the SnV$^-$, with an optional 515~nm charge repump pulse. For SLM excitation, the laser's polarization is optimized with a half-waveplate (HWP) and quarter-waveplate (QWP) before reaching the polarized beam splitter (PBS). A Faraday rotator alters the laser polarization, allowing the SLM-modulated laser signal to be reflected into the cryostat by the PBS. A beam expander (BE) enlarges the beam to cover most of the SLM pixels. The excitation laser beams enter the cryostat through a 4f system (Lenses L1 and L2) after passing through a galvo, which maps the objective back aperture and galvo plane. An HWP and QWP are positioned just before the cryostat window to adjust the setup for cross-polarization or co-polarization collection. The SLM enables the optical manipulation of multiple emitters simultaneously, generating multiple beams on the device to initialize and read out spins within the FOV.


The optical setup is designed to confocally collect photons from the quantum emitter, with the signal passing through a PBS. Situated between two lenses (L3 and L4) in the image plane, a movable D-shaped mirror reflects a portion of light in the image plane (depending on the mirror's location) to another optical path, which is then followed by the camera, galvo, and single-mode fiber collection. A scientific CMOS camera monitors the reflected sub-FOV. Replacing the EMCCD with a single-photon avalanche diode (SPAD) camera that maps the emitter to the pixel enables parallel readout of the spin state in the pulse sequence. By adjusting the HWP, the collection signal can be directed either to the camera path or the confocal collection path, which can enter either the free-space APD or fiber-coupled APD for collection via a programmable flippable mirror. The fiber-coupled signal can be further processed using photonic routers such as optical switches based on microelectromechanical (MEMS) systems, fiber-coupled photonics integrated circuits~\cite{starling2023fully}, or Mach-Zehnder modulators, which can reconfigure the photon interaction among various fiber array channels. This setup enables simultaneous single-mode collection in different sub-FOVs for interference measurement.



\begin{figure*}[ht]%
\centering
\includegraphics[width=1\textwidth]{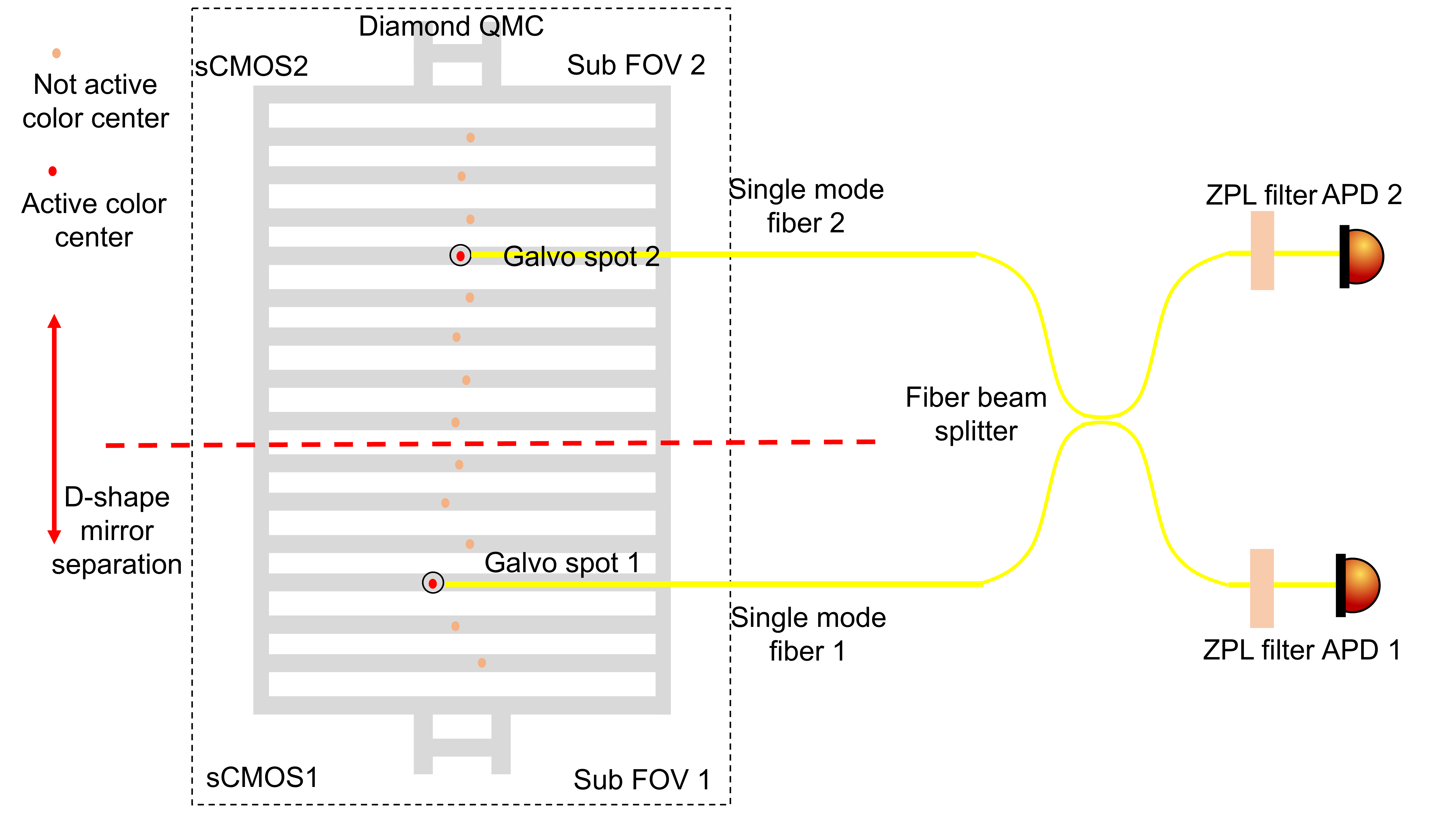}
\caption{\textbf{Approach of the all-to-all routing of photons from quantum emitters.} The D-shaped mirror divides the FOV into two sub-FOVs, each monitored with a scientific CMOS camera. The galvo positions are calibrated with potential color center candidates. Photons from each sub-FOV can be collected into single-mode fibers using corresponding galvos. The collected photons then pass through a fiber beam splitter, followed by a ZPL filter and APD for interference measurement.}\label{figS10}
\end{figure*}

Various methods can be used for exciting and routing photons to enable optical interactions between different quantum emitters~\cite{martinez2022photonic}. For excitation, we use an SLM to generate a programmable laser spot array hologram on the device~\cite{kim2019large}. We then employ a D-shaped mirror, as shown in Fig.\ref{figS9}, to divide the entire free space FOV into multiple sub-FOVs. In each subFOV, we apply a galvo or MEMS mirror to map a single-mode fiber to any position within the subFOV. Once the spin-entangled photons are collected into a fiber, we can perform arbitrary operations for interference and routing using fiber optics. An example of collecting two spin-entangled photons into two sub-FOVs is shown in Fig.\ref{figS10}. The movable D-shaped mirror can divide a large FOV into two subFOVs, enabling interference between any two quantum emitters in different FOVs through time-multiplexing operations. The number of sub-FOVs can be increased with additional D-shaped mirrors, while the number of collection spots within each sub-FOV can be further improved using a single-mode fiber array or a multicore single-mode fiber array.

\end{document}